\documentclass[11pt,fleqn,twocolumn]{article}

\usepackage[a4paper,margin=2cm]{geometry}

\usepackage{amsmath,amssymb}

\usepackage{graphicx}
\usepackage{float}
\usepackage{subcaption}

\usepackage{siunitx}

\usepackage[ruled,vlined]{algorithm2e}

\usepackage[numbers,sort&compress]{natbib}

\usepackage[colorlinks=true, allcolors=blue]{hyperref}

\usepackage{accents}
\newcommand{\bigcdot}{\mathbin{\vcenter{\hbox{\scalebox{1.5}{$\cdot$}}}}}
\newcommand{\widemomentum}[1]{\accentset{\rule{0.7em}{0.4pt}}{#1}}

\title{Performance Benchmarking of Tensor Trains for Quantum-Inspired Homogenization on TPU, GPU, and CPU Architectures}

\author{
Sascha H. Hauck\thanks{Corresponding author: sascha.hannes.hauck@itwm.fraunhofer.de}\\
Fraunhofer ITWM, Kaiserslautern, Germany
\and
Matthias Kabel\\
Fraunhofer ITWM, Kaiserslautern, Germany
\and
Nicolas R. Gauger\\
RPTU Kaiserslautern, Germany
}

\date{}

\begin{document}

\twocolumn[
\maketitle

\begin{abstract}
Recent advances in high-resolution CT-imaging technology are creating a new class of ultra-high resolved microstructural datasets that challenge the limits of traditional homogenization approaches. While state-of-the-art FFT-based homogenization techniques remain effective for moderate datasets, their memory footprint and computational cost grow rapidly with increasing resolution, making them progressively inefficient for industrial-scale problems. To address these challenges, the recently developed Superfast-Fourier Transform (SFFT)-based homogenization algorithm leverages the memory-efficient low-rank representations of Tensor Trains (TTs), which reduce the storage and computational requirements of large-scale homogenization problems. Developed for CPU usage, SFFT-based Homogenization efficiently handles high-resolution datasets, assuming the underlying data is well-behaved.

In this work, we investigate the performance of fundamental TT operations on modern hardware accelerators using the JAX framework. A benchmarking study across CPUs, GPUs, and TPUs evaluates execution times and computational efficiency, highlighting the strengths and limitations of TT operations on different architectures and motivating future hybrid approaches.
Building on these insights, we adapt the SFFT-based homogenization algorithm for accelerator execution, enabling homogenization at high resolutions ranging from 300 million to 70 billion grid points, which are infeasible for the best available GPU-based FFT reference implementation.
While the observed scaling behavior is geometry-dependent, the results demonstrate the potential of accelerator-based quantum-inspired homogenization for high-performance multiscale simulations.
\end{abstract}


\vspace{2em}
]

\section{Introduction}

Homogenization theory provides a framework for predicting the macroscopic behavior of materials with complex microstructures, including composites, polycrystals, and fiber-reinforced alloys \cite{mei96, jablonski16, wang22, huang23, duan22, bakhalov89, zhang23, wang23, gao23}. Central to this approach is the cell problem, a partial differential equation defined over a representative volume element (RVE) that captures the essential features of the microstructure. Solving the cell problem allows one to determine the effective constitutive behavior of the material. For most practical materials, however, analytical solutions are not feasible, necessitating computational techniques \cite{hubbard75, cior99}.

Among these techniques, FFT-based homogenization has emerged as an efficient alternative to classical finite element methods (FEM). Introduced by Moulinec and Suquet  in 1998 \cite{moulinec98}, this approach reformulates the cell problem as a Lippmann-Schwinger equation, which can be solved iteratively using the Fast Fourier Transform (FFT). By discretizing the microstructure on a regular grid, FFT-based methods avoid the challenges of meshing and stiffness matrix assembly inherent to FEM, achieving rapid and accurate solutions for periodic microstructures \cite{toulemonde08}.

Recent advances in CT-imaging technologies now enable the acquisition of microstructural datasets of unprecedented size, reaching tens of billions of voxels \cite{salamon25}. While such high-resolution data provide rich insights, they also pose a significant computational challenge for classical FFT-based homogenization, as times per iteration increase sharply with finer discretization in $d$-dimensional space according to the time-complexity of $\mathcal{O}(dN^d\log N)$ dictated by the FFT \cite{cooley69, moulinec98}.

A straightforward solution is the use of hardware accelerators. However, while the computational load makes GPUs and TPUs essential for large-scale homogenization, the extreme memory requirements of high-resolution data often exceed the capacity of modern accelerators. Efficient GPU-based FFT libraries exist \cite{frigo05}, but FFT-based homogenization algorithms on GPUs typically target small-scale problems or rely on batching strategies for higher-resolution data, providing only moderate speed-ups \cite{leuschner18, kulkarni18, li25, bluehdorn22}. For TPUs, a complete FFT-based homogenization implementation is still missing, although initial FFT implementations show promising performance \cite{lu21}.

Alternative approaches, such as leveraging the Quantum Fourier Transform (QFT) on quantum hardware, could provide exponential speed-ups. However, these applications remain theoretical, as fault-tolerant quantum computers are required even for modest system sizes \cite{givois22, liu24}.

To overcome the computational and memory limitations of classical FFT-based homogenization, a Tensor Train (TT)–based homogenization algorithm has recently been proposed. The Superfast Fourier Transform (SFFT)–based homogenization algorithm \cite{hauck26} employs the TT version of the QFT, breaking the exponential scaling barrier of the FFT while dramatically reducing memory requirements.
However, because this approach is still in an early stage of development, its applicability remains limited to comparatively simple underlying geometries.

In this paper, we make two primary contributions. First, we develop efficient hardware-accelerated implementations of the fundamental TT algebra on modern hardware accelerators, specifically GPUs and TPUs, using the JAX framework. 
We provide a rigorous evaluation of the feasibility and performance of these accelerated TT operations, highlighting the advantages and limitations of each platform, and report detailed benchmarking metrics, including speed and scalability. 
Although libraries for GPU-accelerated TT algebra are already available \cite{cutensornet24}, systematic benchmarking on TPU hardware has been lacking; this work presents the first such evaluation and includes a direct comparison between GPU and TPU performance, building on prior results that demonstrated the suitability of TPUs for the DMRG algorithm \cite{ganahl23}, used to solve optimization problems in TT format.
Additionally, we provide the first practical implementation and systematic evaluation of Polar-based TT compression on modern hardware accelerators \cite{ganahl23}, needed for accurate low-precision arithmetics.

Second, we apply the previously developed hardware-accelerated TT operations to the existing SFFT-based homogenization workflow in a high-resolution setting on GPUs and TPUs.
Our analysis addresses not only the feasibility of running TT-based algorithms on GPUs and TPUs, but also a direct comparison of their performance in terms of computational efficiency and scalability. Importantly, the use of Tensor Trains now enables homogenization of massive datasets, previously infeasible on conventional accelerators, highlighting the potential of low-rank tensor techniques to overcome both memory and computational bottlenecks in industrial-scale simulations.

Together, these contributions establish a foundation for high-performance, large-scale tensor-based homogenization on modern accelerators, demonstrating both the practical viability and performance benefits of TT-based approaches for high-resolution data. Importantly, the present work does not modify the fundamental underlying SFFT-based homogenization algorithm itself, but focuses on its efficient implementation and acceleration using Tensor Train techniques on modern hardware.

The remainder of this paper is organized as follows. We begin with an introduction to Tensor Trains and their fundamental algebra in Section \ref{sec:tensor_trains}. This is followed by a review of state-of-the-art FFT-based homogenization and the recently developed SFFT-based algorithm in Section \ref{sec:fourier_based_homogenization}. Section \ref{sec:benchmark_fundamental} presents a benchmark of the fundamental TT operations in terms of a rigorous performance analysis. The adaptation and application of the SFFT-based homogenization algorithm to GPUs and TPUs is discussed in Section \ref{sec:acceleratedQuantumInspiredHomogenization}. Finally, Section \ref{sec:conclusion} provides a summary and highlights the main conclusions drawn.


\section{Tensor Trains}
\label{sec:tensor_trains}


Emerging independently within mathematics and physics, Tensor Networks (TNs) have evolved into a rich and versatile framework for representing and manipulating high-dimensional data \cite{fannes92, hackbusch09, hackbusch15}.
At their core, TNs express a high-order tensor as a network of interconnected lower-order tensors, linked through prescribed contraction patterns. When the underlying data exhibit suitable structure, such representations can dramatically reduce computational complexity and enable efficient algorithms for large-scale problems.


This work focuses on the Tensor Train (TT) format
\cite{montangere18, biamonte23}, which can be viewed as a
particularly simple yet powerful instance of a general tensor
network. In the TT representation, a high-order tensor
$\mathcal{T}$ is factorized into a sequence of interconnected
third-order tensors according to
\begin{align}
\label{eq:tt_format}
    \mathcal{T}(i_1, ..., i_n) &= \sum_{\boldsymbol{\alpha}}
    A^{\alpha_1}_{\alpha_0}(i_1)
    A^{\alpha_2}_{\alpha_1}(i_2)
    \, ... \,
    A^{\alpha_n}_{\alpha_{n-1}}(i_n)
    \nonumber\\
    &=
    A_1(i_1)
    A_2(i_2)
    \, ... \,
    A_n(i_n), 
\end{align}
where $\boldsymbol{\alpha}$ denotes the collection $(\alpha_0, \alpha_1, \cdots, \alpha_{n-1}, \alpha_n)$.
Each physical index $i_k$ runs from $0$ to $p-1$, however, typically the values of $p$ are the same for all physicl indices.The novel low-order tensors $A_k$ are called the \textit{cores} of the TT and exhibit the shape $(r_k\times p\times r_{k-1})$.
This highlights that the expressive
power of the decomposition depends critically on the
dimensions of the intermediate virtual spaces.
The largest such dimension, known as the TT-rank $r$,
plays a central role as it governs both the representational
capacity of the model and the computational cost of
applied operations.

To illustrate the benefits of the TT formulation, first consider an $n$-th order tensor whose modes each have a (physical) dimension $p$. A naive storage scheme would require $O(p^n)$ memory, which grows exponentially with $n$ and rapidly
becomes infeasible even for moderate dimensions.
In contrast, representing the same tensor in the TT format leads to an exponential reduction  amounting to a memory need of only $O(npr^2)$.
However, a key caveat of this approach is that this method introduces a novel dependence on the TT rank $r$.
This reflects the need for an inherent structure in the data needed to obtain a small rank. 
Besides these memory improvements, their well-established mathematical foundation enables highly efficient computations, often yielding substantial speed-ups while keeping the tensor in its compressed form.

Building on this framework, the Tensor-Train Operator (TTO) extends the TT representation of Equation~\eqref{eq:tt_format} by introducing an additional physical index to each core to encode operator-valued data
\begin{align}
\label{eq:tto_format}
    \hat{\mathcal{T}}(i_1, j_1, \ldots, i_n, j_n) 
    &= \sum_{\boldsymbol{\tilde{\alpha}}}
       \hat{G}_{\tilde{\alpha}_0}^{\tilde{\alpha}_1}(i_1, j_1) 
       ...
       \hat{G}^{\tilde{\alpha}_n}_{\tilde{\alpha}_{n-1}}(i_n, j_n) \nonumber \\
    &= \hat{G}_1(i_1, j_1) ... \hat{G}_n(i_n, j_n),
\end{align}
where $\boldsymbol{\tilde{\alpha}}$ denotes the collection $(\tilde{\alpha}_0, \tilde{\alpha}_1, \cdots, \tilde{\alpha}_{n-1}, \tilde{\alpha}_n)$.
In this representation, each core’s physical index pair $(i_k, j_k)$ corresponds to the incoming and outgoing dimensions of the operator.  
Applying a TTO $\hat{A}$ to a TT $x$, i.e., performing the contraction $y = \hat{A} x$, is analogous to a matrix-vector multiplication, with the TTO acting as a linear operator mapping one TT to another.  
Consequently, this structure forms a cornerstone of the TT formalism and underpins many standard algorithms for efficient data manipulation.


\subsection{Fundamental Algebra}
\label{subsec:tt_algebra}

The most commonly used operations on TTs include addition, multiplication, and linear transformations \cite{oseledets11}. To provide a clear understanding of these operations, we first define the following TTs and TTO
\begin{align}
\label{eq:TTsandTTO}
    x(i_1, \ldots, i_n) &= A_1(i_1) \cdots A_n(i_n), \\
    y(i_1, \ldots, i_n) &= B_1(i_1) \cdots B_n(i_n), \\
    z(i_1, \ldots, i_n) &= C_1(i_1) \cdots C_n(i_n), \\
    \hat{A}(i_1, j_1 \ldots, i_n, j_n) &= \hat{D}_1(i_1, j_1)  \cdots \hat{D}_n(i_n, j_n).
\end{align}

\textit{Constant multiplication}, i.e. multiplying a TT $x$ by a scalar $c$, is the simplest rudimentary operation.  
This can be readly implemented by a simple update rule for the first core 
\begin{align}
\label{eq:tt_const_multiplication}
    A_1(i_1) \rightarrow \tilde{A}_1(i_1)=c \, A_1(i_1).
\end{align}
Since the scalar only rescales a single core, 
the TT rank remain unchanged.

\textit{Addition} of two TTs, $z = x + y$, is performed via diagonalization. 
Specifically, the cores per physical index of the individual TTs are placed along the diagonal of a matrix to form the new cores
\begin{align*}
    C_k(i_k) = 
    \begin{pmatrix}
    A_k(i_k) & 0 \\
    0        & B_k(i_k)
    \end{pmatrix}, \quad k = 2, \ldots, n-1,
\end{align*}
for all internal cores. 
The first and last cores require special treatment due to one of their virtual dimensions being trivial.  
They can be built via a small adaptation of the above formula
\begin{align*}
    C_1(i_1) = 
    \begin{pmatrix}
    A_1(i_1) & B_1(i_1)
    \end{pmatrix}, \hspace{0.5cm}
    C_n(i_n) = 
    \begin{pmatrix}
    A_n(i_n) \\
    B_n(i_n)
    \end{pmatrix}.
\end{align*}
The correctness of this construction follows directly from multiplying the corresponding TT cores of $z$.
Because the block-diagonal construction stacks the cores of $x$ and $y$ along the virtual dimension, the TT ranks of $z$ equal the sum of the corresponding ranks of the two summands.

\textit{Subtraction} of two TTs, $z = x - y$, forms the next complex operation and is implemented in two steps.  
First, the second TT $y$ is multiplied by the scalar $(-1)$ using the update rule for constant multiplication in Equation~\eqref{eq:tt_const_multiplication}, yielding $\tilde{y}$.  
The result can now be obtained in a second step by performing standard addition $z = x + \tilde{y}$.
Since rank growth arises solely from the addition step, the rank of the resulting TT scales as the sum of the ranks of the original TTs.

\textit{Tensor multiplication}, also known as the Hadamard product, between two TTs, $z = x \otimes y$, can be carried out by reordering the operation and the corresponding cores, yielding
\begin{align*}
\label{eq:tt_hadamard}
    z &= \left[A_1(i_1) \cdots A_n(i_n)\right] \otimes \left[B_1(i_1) \cdots B_n(i_n)\right] \\
      &= \left[A_1(i_1) \otimes B_1(i_1)\right] \cdots \left[A_n(i_n) \otimes B_n(i_n)\right].
\end{align*}
Therefore, the cores of $z$ can be identified as
\begin{align*}
    C_k(i_k) = A_k(i_k) \otimes B_k(i_k), \quad k = 1, \ldots, n.
\end{align*}
Regarding the final rank, it becomes clear that it is equal to the product of the ranks of the multiplicands.

\textit{Contraction} between a TT and a TTO, $y = \hat{A} x$, can be performed by summing over the connected physical indices
\begin{align*}
    y_{i} &\equiv y(i_1, \ldots, i_n) \\
    &=
    \sum_{\mathbf{j}}
    \hat{D}_1(i_1, j_1) \cdots \hat{D}_n(i_n, j_n) A_1(j_1)  \cdots A_n(j_n) \\
    &= \sum_{j_1}\hat{D}_1(i_1, j_1) A_1(j_1) \cdots \sum_{j_n}\hat{D}_n(i_n, j_n) A_n(j_n),
\end{align*}
where $\mathbf{j}$ denotes the collection $(j_1, \cdots, j_n)$.
The newly formed cores can be expressed as
\begin{align}
    B_k(i_k) = \sum_{j_k} \hat{D}_k(i_k, j_k) A_k(j_k), \quad k = 1, \ldots, n.
\end{align}
Since the virtual dimensions are not modified during the contraction, their corresponding spaces combine, causing the resulting rank to be equal to the product of the ranks of the involved TT and TTO.
It is noteworthy that the contraction patterns in the TT format are quite simple. However, for moderately more complex TN structures, such as Projected Entangled Pair States (PEPS), the contraction problem is already \#P-complete \cite{schuch07}.

If the result of the TT-TTO contraction is expected to admit a low-rank structure, one can avoid the standard post-compression step by employing a variational optimization strategy. In the Alternating Minimal Energy (AMEn) algorithm, denoted by $\mathcal{A}(\hat{A}, x)=y$, neighboring cores are iteratively contracted and locally optimized, while the representation is enriched using an approximation of the residual $z \approx \hat{A}x - y$ \cite{dolgov13}. This procedure sweeps through all adjacent TT core pairs until convergence is reached, thereby constructing an improved low-rank approximation without explicitly forming large intermediate tensors.
In this way, AMEn yields a quasi-optimal low-rank representation directly in the tensor train manifold, bypassing the need for repeated global compression steps. The method constitutes an extension of the well-known Density Matrix Renormalization Group (DMRG) algorithm with improved convergence properties.

For completeness, we conclude this section by noting that alternative implementations exist for TT operations.  
In particular, it can be advantageous to convert a TT into a TTO. This transformation can improve computational performance, especially when combined with optimization-based methods.  
Assume we have a TT $y$. It can be lifted to a TTO $\hat{A}$, following the format of Equation~\eqref{eq:tto_format}, by introducing delta functions:
\begin{align}
\label{eq:tt_to_tto}
\hat{D}^{\alpha_k}_{\alpha_{k-1}}(i_k, j_k)
=
B^{\alpha_k}_{\alpha_{k-1}}(i_k) \, \delta(i_k, j_k).
\end{align}
The resulting TTO cores $\hat{D}^{\alpha_k}_{\alpha_{k-1}}(i_k, j_k)$ are sparse fourth-order tensors that can be interpreted as the tensor equivalent of a diagonal matrix.
Applying the converted TTO defined in Equation~\eqref{eq:tt_to_tto} to a TT $x$ produces the same result as the Hadamard product of $x$ and $y$.  
Moreover, by leveraging optimization-based methods, the rank of the resulting TT exhibits improved scalability, as it is not fixed to the product of the ranks of $x$ and $y$, but is only upper-bounded by it \cite{hauck26}.

\subsection{Stability}

The operations introduced in the previous section share a common feature: they generally increase the rank of the resulting TT, with the sole exception of trivial scalar multiplication.  
In practical algorithms, however, uncontrolled rank growth over successive operations can lead to numerical instability and rapidly negate the computational advantages of TN methods, resulting in prohibitive runtimes and excessive memory consumption.  
Before addressing strategies for mitigating rank inflation, we first discuss structural conditions that play a key role in maintaining numerical stability.

\textit{Orthogonalization} of a TT is used to enforce additional orthogonality conditions on its component cores. 
This step is required not only as a precursor to compression algorithms but also to improve the overall numerical stability of the TT formalism. 
Let us consider a TT $x$ defined according to Equation~\eqref{eq:TTsandTTO}, whose cores do not initially satisfy any such criteria. 
The aim of orthogonalization is to restructure the underlying cores such that the TT assumes the form
\begin{align*}
    x_i &\equiv x(i_1, \ldots, i_n) \\
    &= U_1(i_1) \cdots U_{m-1}(i_{m-1}) T(i_m) \\
    & \times V_{m+1}(i_{m+1})\cdots V_n(i_n),
\end{align*}
such that the left-isometry conditions
\begin{align*}
    \sum_{i_k}
    \sum_{\alpha_{k-1}} U_{\alpha_{k-1}}^{\alpha_k}(i_k)
    {U^{\dagger}}^{\alpha_k'}_{\alpha_{k-1}}(i_k)
    =
    \delta_{\alpha_k, \alpha_{k}'}
\end{align*}
and the right-isometry conditions
\begin{align*}
    \sum_{i_k}
    \sum_{\alpha_k}
 {V^{\dagger}}^{\alpha_k}_{\alpha_{k-1}}(i_k)
    V_{\alpha_{k-1}'}^{\alpha_k}(i_k)
    =
    \delta_{\alpha_{k-1}, \alpha_{k-1}'}
\end{align*}
are satisfied. 
The remaining tensor $T(i_m)$ is called the orthogonality center and may be interpreted as the non-orthogonal residuum of the TT. This representation is commonly referred to as the canonical form of the TT $x$. 

For simplicity, we restrict ourselves to the case $m\in\{1,n\}$, meaning that the orthogonality center is located at one of the boundary cores. 
This canonical form can be constructed by applying a QR decomposition from left to right (or, equivalently, an RQ decomposition from right to left) to each core of the TT. In this process, the orthogonal matrix $Q$ is retained, while the corresponding rectangular matrix $R$ is absorbed into the next core, after which the procedure continues in the same manner:
\begin{align*}
        x_i &= A_1(i_1) \, A_2(i_2)\cdots A_{n-1}(i_{n-1}) A_n(i_n), \\
        &= Q_1(i_1) \, \left[R_1\, A_2(i_2)\right] \cdots A_{n-1}(i_{n-1}) A_n(i_n), \\
    &= Q_1(i_1) \, Q_2(i_2)\, R_2 \cdots A_{n-1}(i_{n-1})A_n(i_n), \\
    & \cdots\\
    &= Q_1(i_1) \, Q_2(i_2) \cdots Q_{n-1}(i_{n-1}) \left[R_{n-1} A_n(i_n)\right], \\
    &= Q_1(i_1) \, Q_2(i_2) \cdots Q_{n-1}(i_{n-1}) T(i_n).
\end{align*}
Since the rectangular matrix in each step is in fact square, the TT ranks of the individual cores remain unchanged throughout the orthogonalization procedure.


\textit{Compression} addresses the issue of rank growth that accumulates over successive operations and is therefore one of the most essential procedures in TN algorithms. Without compression, the ranks of intermediate TTs would quickly become prohibitively large, making computations infeasible. 


Let us first discuss \textit{Compression via SVD}.  
We begin by left-to-right orthogonalization of TT $x$.  
A Singular Value Decomposition is then applied to the orthogonality center (i.e., the last core):
\begin{align*}
    T(i_n) &\rightarrow Q_n \mathrm{diag}(\sigma_1, \sigma_2, ..., \sigma_l, ...,\sigma_q) V_n^\dagger(i_n)\\
    &\approx Q_n \mathrm{diag}(\sigma_1, \sigma_2, ..., \sigma_l) V_n^\dagger(i_n) \\
    &= \tilde{Q}_n V_n^\dagger(i_n).
\end{align*}
Here, the smallest $q-l$ singular values are truncated.  
Since the singular values are ordered in descending magnitude, this truncation provides a controlled reduction of the virtual dimension and thus limits the TT ranks.
The truncated singular matrix is then absorbed into $\tilde{Q}_n$, which is subsequently multiplied into the next core to the left, while $V_n^\dagger(i_n)$ remains as the newly updated core at the $n$-th position.
This shifts the orthogonality center one position to the left, after which the same procedure is applied again.  
The process continues until the entire TT has been traversed.

Finding an appropriate number of singular values to retain can be approached in two general ways. 
The simple option is to prescribe a fixed number of singular values to keep. 
However, this strategy provides no guarantee that the truncated result remains a sufficiently accurate approximation of the original TT.
A more reliable alternative is to determine the truncation adaptively such that the approximation $\tilde{x}$ of a TT $x$ satisfies
\begin{align}
    \label{eq:rounding_condition_tt}
    ||x-\tilde{x}||_F \leq \varepsilon ||x||_F,
\end{align}
where $||\cdot||_F$ denotes the Frobenus norm.
A sufficient condition for Equation~\eqref{eq:rounding_condition_tt} is that the root of the squared sum over the smallest singular values $\sigma_k$ does not exceed
\begin{align}
    \label{eq:optimal_delta}
    \delta = \frac{\varepsilon}{\sqrt{1-n}} ||x||_F,
\end{align}
which ensures that only the significant singular values are retained \cite{oseledets11}.


\textit{Compression via Polar Decomposition} provides an alternative compression approach that is less sensitive to the conditioning of the input. As a result, it is particularly well-suited for systems with low-precision arithmetic.
Instead of performing the SVD, the polar decomposition is employed on the orthogonality center
\begin{align*}
    T(i_n) = U \, H,
\end{align*}
where $U$ is isometric and $H$ is positive semidefinite, with the singular values $\sigma_k$ being intrinsic to $H$.

To extract the components of $W$ associated with singular values above a certain $\tilde{\delta}$, we consider the Hermitian matrix $H - \tilde{\delta} I$ and perform its polar decomposition
\begin{align*}
    H - \tilde{\delta} I = U' \, H'.
\end{align*}
The reasoning behind this being that the now newly formed unitary factor $U'$ has eigenvalues $(+1)$ for eigenvectors corresponding to $\sigma_k > \tilde{\delta}$ and $(-1)$ for $\sigma_k \leq \tilde{\delta}$.
Subsequently, projectors onto the "above cutoff" and "below cutoff" subspaces are constructed as
\begin{align*}
    P_{\pm} = \frac{1 \pm U'}{2}.
\end{align*}
Through these, the truncated part of $H$ can readily be obtained by projection onto the "above cutoff" subspace
\begin{align*}
    H_{>\tilde{\delta}} = P_{+} \, H \, P_{+},
\end{align*}
yielding the reduced orthogonality center
\begin{align*}
    \tilde{T}(i_n) = U \, H_{>\tilde{\delta}}.
\end{align*}

At this stage, $\tilde{T}(i_n)$ still retains the original number of columns and thus has no real compression yet. To construct real truncated factors $F$ and $G$ (with $l<q$ columns and rows, respectively) such that $\tilde{T}(i_n) = F G$, an orthonormal basis $C_{+}$ for the column space of $P_{+}$ is computed via subspace iteration. The resulting factors can be obtained through
\begin{align*}
    F = U \, C_{+}, \quad
    G = C_{+}^{\dagger} \, H,
\end{align*}
producing the truncated representation without requiring an explicit SVD.

Comparing the two approaches, SVD-based compression is generally preferred due to its lower computational cost and faster convergence, provided that the SVD is numerically stable. In addition, SVD-based truncation yields a quasi-optimal rank approximation that satisfies inequality~\eqref{eq:rounding_condition_tt} when an appropriate threshold $\delta$ is chosen (e.g., as in equation~\eqref{eq:optimal_delta}). 

By contrast, the polar-based algorithm relies on an absolute cutoff $\tilde{\delta}$, which can be problematic because the singular values themselves are not explicitly available. Nevertheless, by setting
\begin{align*}
    \tilde{\delta} = \frac{\delta}{q},
\end{align*}
where $q$ is the total number of untruncated singular values, inequality~\eqref{eq:rounding_condition_tt} can still be strictly enforced. This  results in more a conservative truncation, producing a TT rank that exceeds the optimal SVD-based result.

A detailed comparison of the run-time performance between SVD- and polar-based compression methods is provided in Appendix~\ref{app:SVD_vs_Polar}.


\section{Fourier-based \\Homogenization}
\label{sec:fourier_based_homogenization}
In recent years, computational homogenization -- particularly Fourier-based methods -- has undergone substantial development \cite{schneider21}. Among the most prominent advances are quantum and TN-based approaches, which either leverage quantum computing paradigms or exploit compressed representations of microstructural fields, enabling significant computational acceleration.

In this work, we focus on the latter, i.e., the quantum-inspired TN-based formulation. We begin by reviewing the classical FFT-based scheme of Moulinec and Suquet (1998) \cite{moulinec98}, which forms the foundation of these developments, before introducing the quantum-inspired extension in detail. For explicit quantum-computing formulations, we refer the interested reader to Liu et al. \cite{liu24}.

\subsection{FFT-based Homogenization}
\label{subsec:fft_based_homogenization}

The primary objective of the FFT-based Homogenization is to determine the effective linearly elastic macroscopic stiffness tensor $\mathbf{C}^\mathrm{eff}$ of an underlying composite, 
assuming the properties of the individual constituents are known.
The stiffness tensor $\mathbf{C}^\mathrm{eff}$ is defined through the macroscopic stress-strain relationship
\begin{align}
\label{eq:macro_ss}
    \mathbf{\sigma}^\mathrm{eff} = \mathbf{C}^\mathrm{eff} \mathbf{E},
\end{align}
where $\mathbf{\sigma}^\mathrm{eff}$ and $\mathbf{E}$ represent the macroscopic stress and strain tensors, respectively.

To solve the above equation, it is advantageous to perform the analysis at the local scale of the heterogeneities in the composite. In this context, a rectangular representative volume element (RVE) $\mathcal{V} = [0, L_1] \times [0, L_2]$ is introduced, which is assumed to be significantly smaller than the characteristic macroscopic length scale \cite{cior99}.
This scale separation allows the macroscopic problem to be reformulated as a microscopic boundary value problem on the RVE. This procedure, in which the global macroscopic response is obtained via a local substitute problem, is commonly referred to as the corrector problem \cite{brisard10}.


In this setting,
assuming periodic boundary conditions on the RVE,
the local strain field can be written as
\begin{align}
\label{eq:local_strain}
    \mathbf{\varepsilon(x)}
    = \mathbf{E} + \nabla^\mathrm{S} \mathbf{u(x)},
\end{align}
where $\nabla^\mathrm{S}$ and $\mathbf{u(x)}$ are the symmetrized nabla-operator and the local displacement fluctuations,
respectively.
By imposing that the gradient of the displacement field vanishes over the RVE, 
\begin{align*}
    \frac{1}{|\mathcal{V}|}\int_\mathcal{V} \nabla\mathbf{u(x)} \mathrm{d}\mathbf{x}=0,
\end{align*}
it is assured that the macroscopic strain $\mathbf{E}$ represents the mean strain over the RVE.
Since the macroscopic strain is assumed to be known,
the focus of the derivation shifts to finding a representation of the fluctuation term $\nabla^\mathrm{S} \mathbf{u(x)}$.

We continue the derivation by identifying that the local stress tensor must satisfy the force-equilibrium equation,
given by
\begin{align}
\label{eq:force_eq}
    \operatorname{div}\mathbf{\sigma(x)} 
    &= \operatorname{div}\left\{\mathbf{C(x)}\left[\mathbf{E} + \nabla^\mathrm{S}\mathbf{u(x)}\right]\right\}    \\
    &= 0.
    \nonumber
\end{align}
To address the local problem, 
equation \eqref{eq:force_eq} is reformulated through an isotropic reference material,
characterized by a constant stiffness tensor $\mathbf{C}_0$ \cite{brisard10},
leading to 
\begin{align}
\label{eq:pol_force_eq}
    \operatorname{div}\left\{\mathbf{C}_0\left[\mathbf{E} + \nabla^\mathrm{S}\mathbf{u(x)}\right] + \tau(\mathbf{x})\right\} = 0.
\end{align}
with the polarization tensor
\begin{align}
\label{eq:polarization_tensor}
    \tau(\mathbf{x}) = \mathbf{\left[C(x)-C_0\right] \left[E + \nabla^\mathrm{S} u(x)\right]}.
\end{align}
The closed-form differential equation \eqref{eq:pol_force_eq} can be solved using a Green's function $G_0$, 
resulting in the following expression for the displacement gradient
\begin{align}
    \label{eq:green_conv}
    \nabla^\mathrm{S} \mathbf{u(x)} = -\left(\varGamma_0 \ast \tau\right)(\mathbf{x}),
\end{align}
where $\varGamma_0= \varepsilon G_0 \operatorname{div}$ denotes the Green-Eshelby operator for strains \cite{grimm21}.
By substituting solution \eqref{eq:green_conv} into the strain expression \eqref{eq:local_strain},
and applying the definition of the polarization tensor \eqref{eq:polarization_tensor},
we obtain a fixed-point formulation of the Lippmann–Schwinger equation
\begin{align}
\label{eq:lippmann_schwinger}
       \varepsilon(\mathbf{x}) = \mathbf{E} - \left\{\varGamma_0 \ast \left[\left(\mathbf{C - C}_0\right)\varepsilon\right]\right\}(\mathbf{x}).
\end{align}

However,
since the Green-Eshelby operator has a known analytical form in Fourier space,
equation \eqref{eq:green_conv} can be transformed accordingly,
yielding
\begin{align}
    \label{eq:green_conv_momentum}
    \mathcal{F}\left[\nabla \mathbf{u}\right](\mathbf{q}) = -\widemomentum{\varGamma}_0(\mathbf{q})\, \widemomentum{\tau}(\mathbf{q}),
\end{align}
where the Fast Fourier Transform $\mathcal{F}[\mathbf{\,\bigcdot\,}]$ is employed for efficient computation of equation \eqref{eq:green_conv} in momentum space.
Throughout this paper, variables in momentum space are denoted by a wide line above the symbol to distinguish them from their counterparts in spatial space.

Finally,
the FFT-based Homogenization algorithm \ref{algo:fft} is derived by 
incorporating the Fourier representation of the Green-Eshelby operator \eqref{eq:green_conv_momentum} into the Lippmann-Schwinger Equation \eqref{eq:lippmann_schwinger}.
The algorithm iteratively refines the initial approximation of the local strain,
conveniently initialized as the mean strain $\mathbf{E}$, by accounting for local variations in the material structure.
 The iterations continue until a predefined accuracy threshold $\delta_\mathrm{tol}$ is reached.

\begin{algorithm}[!h]
    \SetKwInOut{Input}{Input}
    \Input{Mean strain field $\varepsilon_0=\mathbf{E}$}
    \While{not converged}{
        $\tau_m = \left(\mathbf{C} - \mathbf{C}_0\right) \varepsilon_m\;$
        
        $\widemomentum{\tau}_m = \mathcal{F}\left[\tau_m\right]$
        
        $\widemomentum{\varepsilon}_m =\widemomentum{\varGamma}_0\, \widemomentum{\tau}_m$
        
        $\varepsilon_{m+1} = \mathbf{E} -\mathcal{F}^{-1}\left[\widemomentum{\varepsilon}_m\right]$
        
        \If{$\|\frac{\varepsilon_{m+1} - \varepsilon_m}{\mathbf{E}} \| < \delta_\mathrm{tol}$}{
            break;
        }
        $m = m+1$
    }
    
    \textbf{Return}: Local strain field $\varepsilon_\text{local}\equiv\varepsilon_{m+1}$
    
    \caption{FFT-based Homogenization}
    \label{algo:fft}
\end{algorithm}

With the output obtained from algorithm \ref{algo:fft}, we can readily calculate the effective stiffness operator $\mathbf{C}^\mathrm{eff}$ through equation \eqref{eq:macro_ss}.
A detailed explanation of this procedure can be found in Appendix \ref{app:extractingStiffness}.
We conclude this Section by noting that the computational complexity of the FFT-based algorithm \ref{algo:fft} is dominated by the FFT operations,
scaling with $\mathcal{O}\left(d N^d \log N\right)$,
where $N$ is the number of discretization points per dimension $d$.
Thus,
improving the FFT-based Homogenization procedure relies on either enhancing the performance of FFT operations,
such as through hardware acceleration,
or circumventing the classical FFT entirely.

\subsection{Quantum Inspired \\Homogenization}
\label{subsec:quantum_inspired_homogenization}

In the following, we provide a concise overview of the SFFT-based homogenization algorithm, which serves as the foundation for the implementations employed in our experiments. A detailed derivation and formal development of the method can be found in the original paper~\cite{hauck26}.  
For clarity and consistency throughout this work, the terms "SFFT-based algorithm" and "Quantum-Inspired algorithm" will be used interchangeably to refer to the same method.

\begin{figure}
    \centering
    \includegraphics[width=\linewidth]{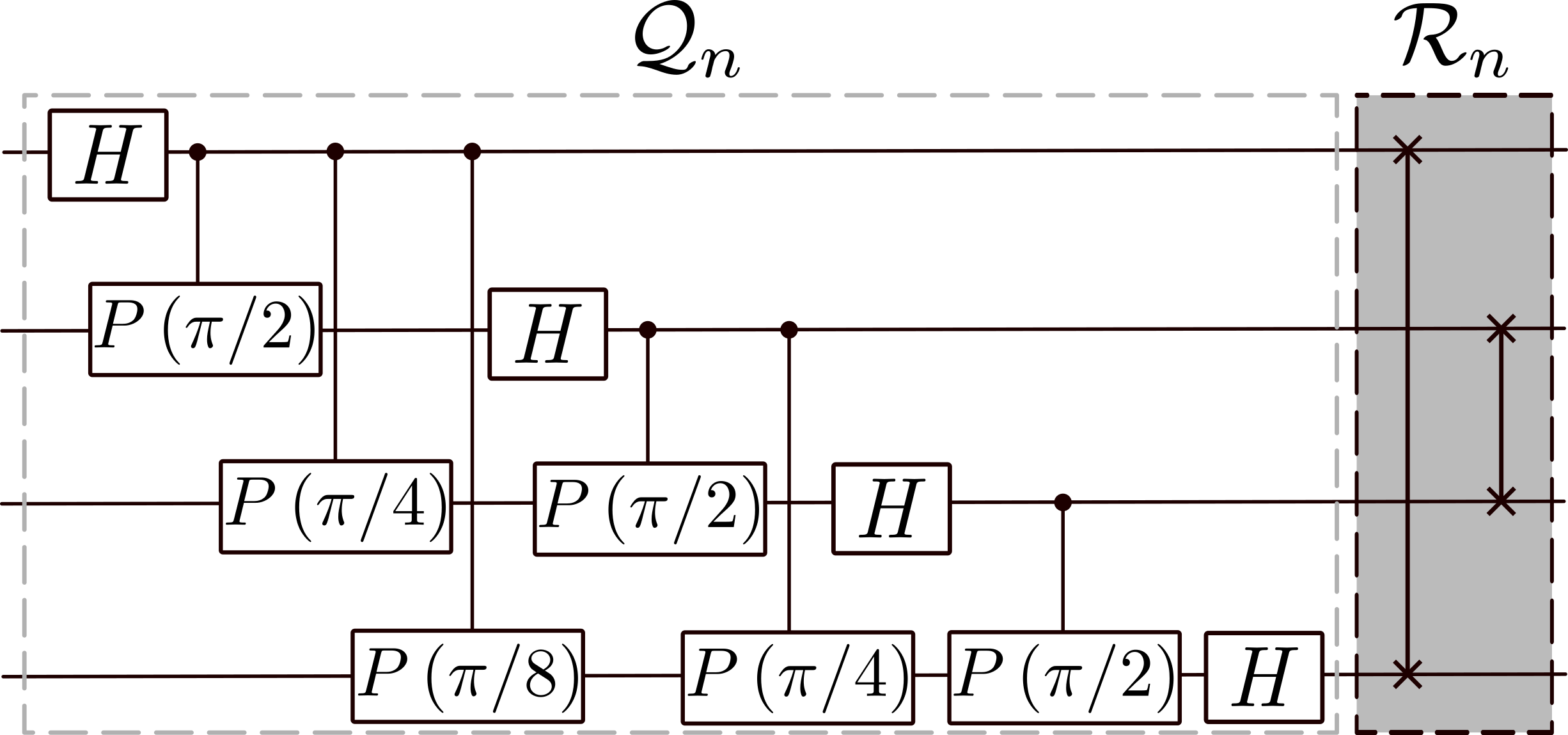}
    \caption{Quantum circuit for the four-qubit Quantum Fourier Transform, $\mathcal{QFT}_4 = \mathcal{R}_4 \mathcal{Q}_4$.}
    \label{fig:qft_qc}
\end{figure}

To construct the SFFT-based Homogenization algorithm, 
the operands and operators used in the classical FFT-based algorithm~\ref{algo:fft} must first be represented in TT and TTO format, respectively.  
A straightforward approach would employ the TT variant of the FFT. While this may reduce memory usage, it does not alleviate the computational bottleneck, as the complexity remains that of the classical FFT.  
We therefore replace the FFT with the $n$-qubit Quantum Fourier Transform (QFT), denoted as $\mathcal{QFT}_n$, to leverage its exponentially improved computational complexity, scaling as $\mathcal{O}(d \log^2 N)$.  
The standard QFT formulation, however, is unsuitable for a TTO representation due to exponential rank growth for increased resolution of the underlying domain.  
Fortunately, this limitation applies only to the standard QFT representation commonly found in the literature.  
By separating the QFT as $\mathcal{QFT}_n = \mathcal{R}_n \, \mathcal{Q}_n$ into a "heavy-lifting" component $\mathcal{Q}_n$ and a reordering contribution $\mathcal{R}_n$ (see Fig.~\ref{fig:qft_qc}), the circuit associated with $\mathcal{Q}_n$ can be represented as a low-rank TTO \cite{chen23}.  
Since the circuit $\mathcal{R}_n$ merely enforces a qubit reordering, it can be safely ignored, provided the new convention is consistently applied in momentum space.
Due to its improved speed, the TTO representation of $\mathcal{Q}_n$ is generally referred to as the "Superfast Fourier Transform" (SFFT), giving the method its name \cite{dolgov12}.
Thus, the algorithm will use both the SFFT $\hat{\mathcal{S}}$ as well as its inverse $\hat{\mathcal{S}}^{-1}$ instead of the regular FFT.
Due to the close connection between the SFFT and quantum circuits, the associated TTO has the same number of cores as the corresponding QFT circuit has qubits. Furthermore, the physical dimension of the operator is set by the qubit dimensionality, i.e., $p=2$.

The remaining operands and oerators -- namely,
the local strain $\varepsilon(\mathbf{x})$, 
the mean strain $\mathbf{E}$,
the Green-Eshelby operator $\mathnormal{\widemomentum{\Gamma}}_0(\mathbf{q})$ and the local and isotropic reference stiffnesses $\mathbf{C(x)}$ and $\mathbf{C}_0$ -
must be transferred into TT format as well. 
This is achieved in two steps:
First, 
operands depending on the continuous variables $\mathbf{x}$ and $\mathbf{q}$ will be discretized in a binary fashion
\begin{align}
\label{eq:binary_discretization}
    t(x) \approx t\left(\sum_{j=1}^n b_j 2^{-j}\right) = \mathcal{T}(b_1 b_2...b_n),
\end{align}
with $b_j\in\{0,1\}$,
resulting in a high-order tensor in multi-index formulation.
The binary-fraction representation is more efficient than a standard discretization $t(x)\approx t[(k-1)\Delta x]=F(k)$ with a single index $k$, as it allows separation into multiple tensor cores, one per binary index, improving TT applicability. Additionally, it naturally yields an $n$-th order tensor with physical index dimension $p=2$, consistent with the SFFT representation.

Once the high-order tensor is obtained,
the cores are separated by applying the TT-Cross-Approximation algorithm \cite{oseledets11, oseledets10}.
In our multidimensional case, dimensions are processed sequentially, with cores grouped by dimension in their TT representations. For instance, in a 2D example with five cores per dimension, the first five cores correspond to the discretization of the $x$ dimension, while the remaining five cores represent the $y$ discretization.
Throughout this paper, we assume $N=2^n$ discretization points per spatial dimension $d$,
corresponding to $n$ cores -- analogous to $n$ qubits -- per dimension.


So far, only the “inner” indices arising from the spatial discretization have been considered. In practice, however, the quantities of interest are tensorial objects defined in physical space. For example, in Voigt notation (see Appendix~\ref{app:extractingStiffness}), strain is represented as a vector and stiffness as a matrix, introducing additional “outer” indices beyond the spatial ones.
To incorporate these, each operand can be treated as a vector or matrix stored in TT format.
While this strategy is valid, greater improvements in both computational speed and memory efficiency are obtained by representing the full operand, including its outer indices, directly in TT form.
A more efficient approach is to include the outer indices directly in the TT representation by introducing an additional control core. This core may have physical dimension different from $p=2$ and acts as a selector for the components of the Voigt representation, encoding its algebraic structure rather than spatial information.
For simplicity, we continue to refer to this extended representation as TT(O).

The final operator that requires special treatment is the Green-Eshelby operator $\widemomentum{\varGamma}_0(\mathbf{q})$.  
Since this operator resides in momentum space, it must be adapted to comply with the momentum space convention imposed by the SFFT.  
To achieve this, a reordering of the TT cores is necessary, which involves two steps:  
\begin{enumerate}
    \item The indices corresponding to each geometrical dimension $d$ of the initial discretization tensor, obtained via Equation~\eqref{eq:binary_discretization}, must be reversed.
    \item The cores need to be complex conjugated.
\end{enumerate}
Applying these additional steps to operators in momentum space ensures consistency with the SFFT convention.  
After translating the original algorithm into TT format, the final result for the local strain field is obtained by contracting the TT into its full data representation.  
This contraction is performed by summing over all virtual dimensions of the TT, as defined in Equation~\eqref{eq:tt_format}.  
Combining all steps results in the final form of the SFFT-based homogenization algorithm, summarized in Algorithm~\ref{algo:sfft}.

\begin{algorithm}[!h]
    \SetKwInOut{Input}{Input}
    \Input{Mean strain field TT $\varepsilon_0=\mathbf{E}$}
    \While{not converged}{
        $\tau_m = \left(\mathbf{C} - \mathbf{C}_0\right)\varepsilon_m\;$
        
        $\widemomentum{\tau}_m = \mathcal{S}\left[\tau_m\right]$
        
        $\widemomentum{\varepsilon}_m = \mathcal{A}(\widemomentum{\varGamma}_0\, , \widemomentum{\tau}_m$)
        
        $\varepsilon_{m+1} = \mathbf{E} - \mathcal{S}^{-1}\left[\widemomentum{\varepsilon}_m\right]$
        
        \If{$\|\frac{\varepsilon_{m+1} - \varepsilon_m}{\mathbf{E}} \|_F < \delta_\mathrm{tol}$}{
            break;
        }
        $m = m+1$

    }
    $\varepsilon_\text{local} = \text{contract}(\varepsilon_{m+1})$ \;

\textbf{Return}: Local strain field $\varepsilon_\text{local}$
    
    \caption{SFFT-based Homogenization}
    \label{algo:sfft}
\end{algorithm}

In terms of computational speed, the SFFT-based algorithm achieves an exponential acceleration with respect to both the number of discretization points $N = 2^n$ and the geometric dimensionality $d$, compared to the classical FFT-based homogenization method, provided that the TT ranks remain manageable \cite{hauck26}.  
This caveat, i.e. the requirement of small ranks, depends crucially on the underlying geometry of the RVE.  
However, the geometry-dependent term can be precomputed during an offline phase, enabling an informed assessment of the algorithm's applicability prior to executing the full computation.

In the original formulation, after obtaining the TT(O) representations of all operands and operators,  
the TTs involved in the Hadamard multiplications within the algorithm were lifted to TTOs  
(see Section~\ref{subsec:tt_algebra}).  
This elevation enabled the use of optimization-based methods, which are computationally advantageous  
in CPU-based environments. In particular, this is beneficial for operations involving the high-rank Green-Eshelby operator, for which the adapted AMEn optimization algorithm is employed \cite{dolgov13}.

\section{Benchmarking Fundamental Operations}
\label{sec:benchmark_fundamental}

\begin{figure*}[tbph]
    \centering
    \includegraphics[width=\textwidth]{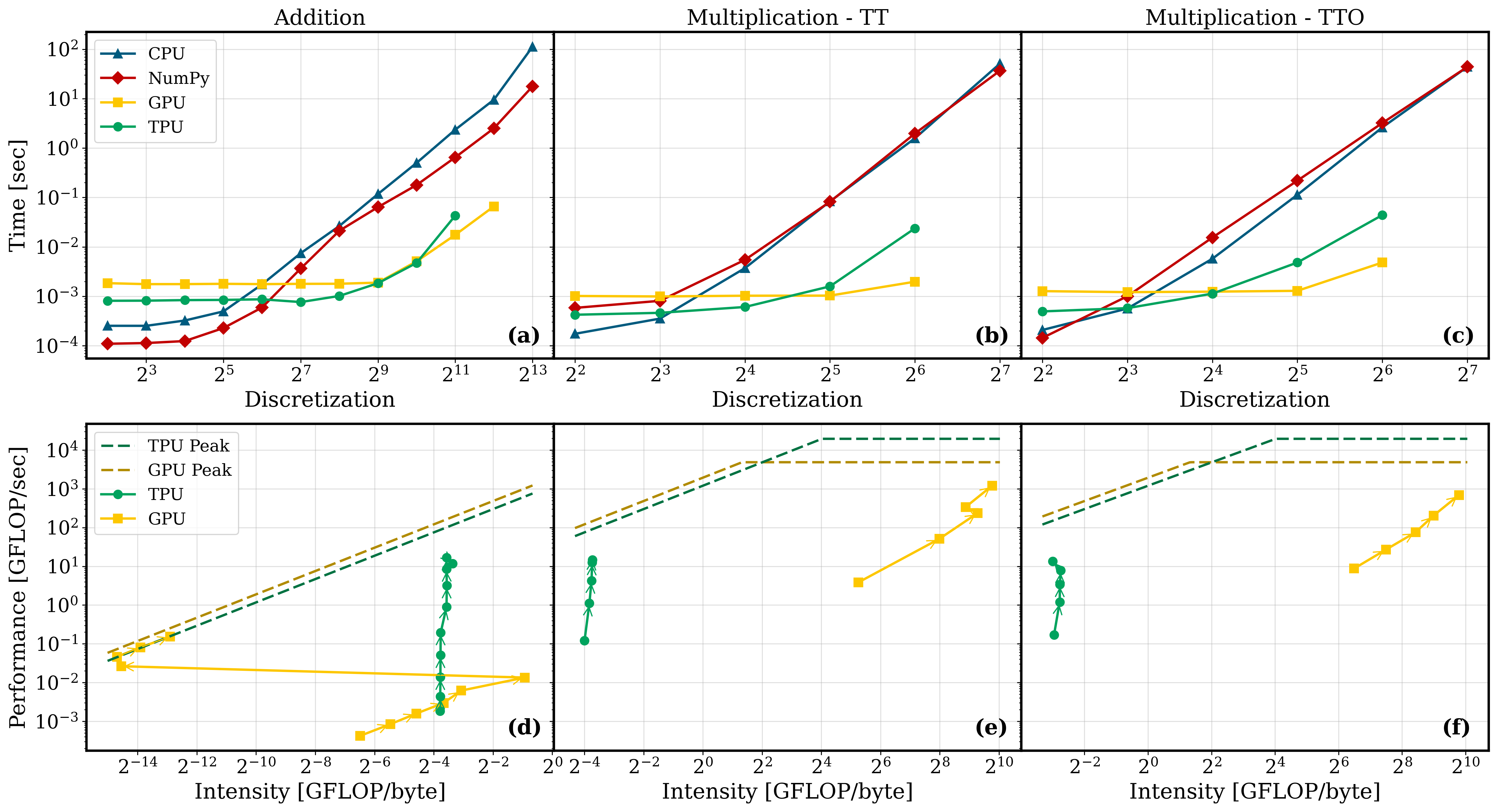}
\caption{Parallel TT operations. Columns show TT addition (a,d), TT multiplication (b,e), and TT–TTO contraction (c,f). The first row shows runtime benchmarks, the second the GPU and TPU Roofline models.}
    \label{fig:parallelizableTTOperations}

\end{figure*}

Recent work in heterogeneous high-performance computing has extensively studied the acceleration of sparse tensor contraction and  Canonical Polyadic Decomposition (CPD) kernels, particularly for irregular sparse tensor–tensor and tensor–matrix operations. These approaches typically focus on load balancing, tiling strategies, and memory optimization under sparsity-induced irregular memory access patterns on CPU–GPU systems \cite{Xiao24, Chen22}.

In contrast, the present work targets structured TT and TTO computations arising in SFFT-based homogenization. Here, the computational graph is deterministic and sparsity is encoded via low-rank tensor-network structure rather than irregular index patterns leading to fundamentally different performance characteristics and, consequently, different optimization trade-offs on heterogeneous hardware.
Even though we investigate TT operations on each of the available platforms (CPU, GPU, TPU) separately, these operations constitute the building blocks of heterogeneous TT-based computing workflows.

For the simulations we restrict ourselves to combinations between three different hardware configurations and two different compute modes. 
Specifically,
we are using an Dual Intel Xeon Gold 6240R CPU from Intel,
an A100 GPU from Nvidia
as well as Googles TPU v4.8. 
Additional Information on the Hardware specifications used can be found in Appendix \ref{app:hardware}.

For the compute modes, we distinguish between two implementations:
(i) An implementation that stores the TT-cores as arrays in a list, either as JAX \cite{jax18} or NumPy arrays \cite{harris20}.
In what follows, the specific array type will be clear from the context.
(ii) An implementation that stores the TT-cores as arrays within another (batched) array.
This second version has the drawback that the cores might need to be padded in order to fit the dimensionality of the batched format.
We will refer to the first mode as "list-format" while the latter one is called "batched-format".
This distinction was made since they tackle different regimes and thus are expected to be the most efficient under different hardware conditions.
While the first version is more memory efficient, the latter is tailored to the specific architecture of Googles TPU.
However, our benchmarks showed that the list format outperforms the batched format for the high memory needs that we are interested in.
To make the upcoming discussion concise, we will restrict ourselves to the discussion about the list-format.
A detailed discussion about the differences as well as additional benchmarks on the batched variant can be found in Appendices \ref{app:jax_tensor_Homogenization} and \ref{app:accuracy}.

\subsection{Software Setup}
\label{subsec:softwaresetup}

To ensure reproducibility and a fair comparison across architectures, we briefly summarize the software stack used for the accelerator implementations.
Google's Tensor Processing Units (TPUs) are programmed through XLA \cite{xla25}, an optimizing compiler that lowers high-level operations (HLOs) to low-level operations (LLOs) executable on the hardware. While HLOs can be written manually, they are typically generated automatically by tracing computations expressed in higher-level programming interfaces—JAX being a prominent example. JAX transforms Python functions into a computation graph and delegates its compilation to XLA, which produces highly optimized GPU and TPU executables.

This compilation is performed just-in-time (JIT). The first invocation of a JIT-compiled function initiates tracing and compilation, making it substantially more expensive than subsequent executions. This initial overhead, commonly referred to as the warm-up phase, can dominate runtime and thereby diminish or even negate the performance benefits of JAX for short-lived programs.

To alleviate this issue, JAX offers persistent compilation, which allows compiled executables to be stored externally and reused across multiple program runs. This mechanism bypasses repeated warm-up overhead and enables stable performance from the first invocation. Since this feature -- although still experimental at the time of writing -- effectively eliminates warm-up costs for repeated executions, we omit warm-up time from the benchmarks presented in this section.




Originally TPUs were developed as linear algebra accelerators with the use case of performing faster neural network inference.
Since speed is of the essence and weights within neural networks generally don't need high accuracy to reach their full potential \cite{romanov21}, 
TPUs are optimized around the fast BF16 standard.
For our analysis this accuracy is not sufficient.
In fact, we will need to set the accuracy to its highest possible value of the complex64 standard in order to get accurate results for our use-case.
Due to our high-accuracy requirements, this analysis deliberately operates TPUs outside their typical regime and should therefore be viewed as complementary to standard TPU use-cases.

To ensure a fair comparison to CPU and GPU,
we will further adapt their accuracy to the complex64 standard as well.
To benchmark the fundamental TT operations,
we used Quantized TTs and TTOs with randomly initialized cores. We have obtained the data-points in our upcoming analysis by taking the mean over 100 successive runs per probed operation.

Iteration times were measured using Python’s timeit library, averaging over 100 runs per operation. Performance on TPUs was obtained by collecting GFLOPS values through the JAX Profiler. For GPU benchmarking, we used the NVIDIA Tools Extension (NVTX), since in this context the JAX Profiler does not correctly account for computations executed in precompiled CUDA libraries, leading to significantly underestimated GFLOPS measurements.

For both accelerators, certain data points in the high-discretization regime could not be obtained due to out-of-memory (OOM) errors. In the performance benchmarks, OOM conditions arise even at lower discretization levels, because additional figures of merit must be tracked, further reducing the memory available for computation.

\begin{figure*}[tbph]
    \centering
    \includegraphics[width=\textwidth]{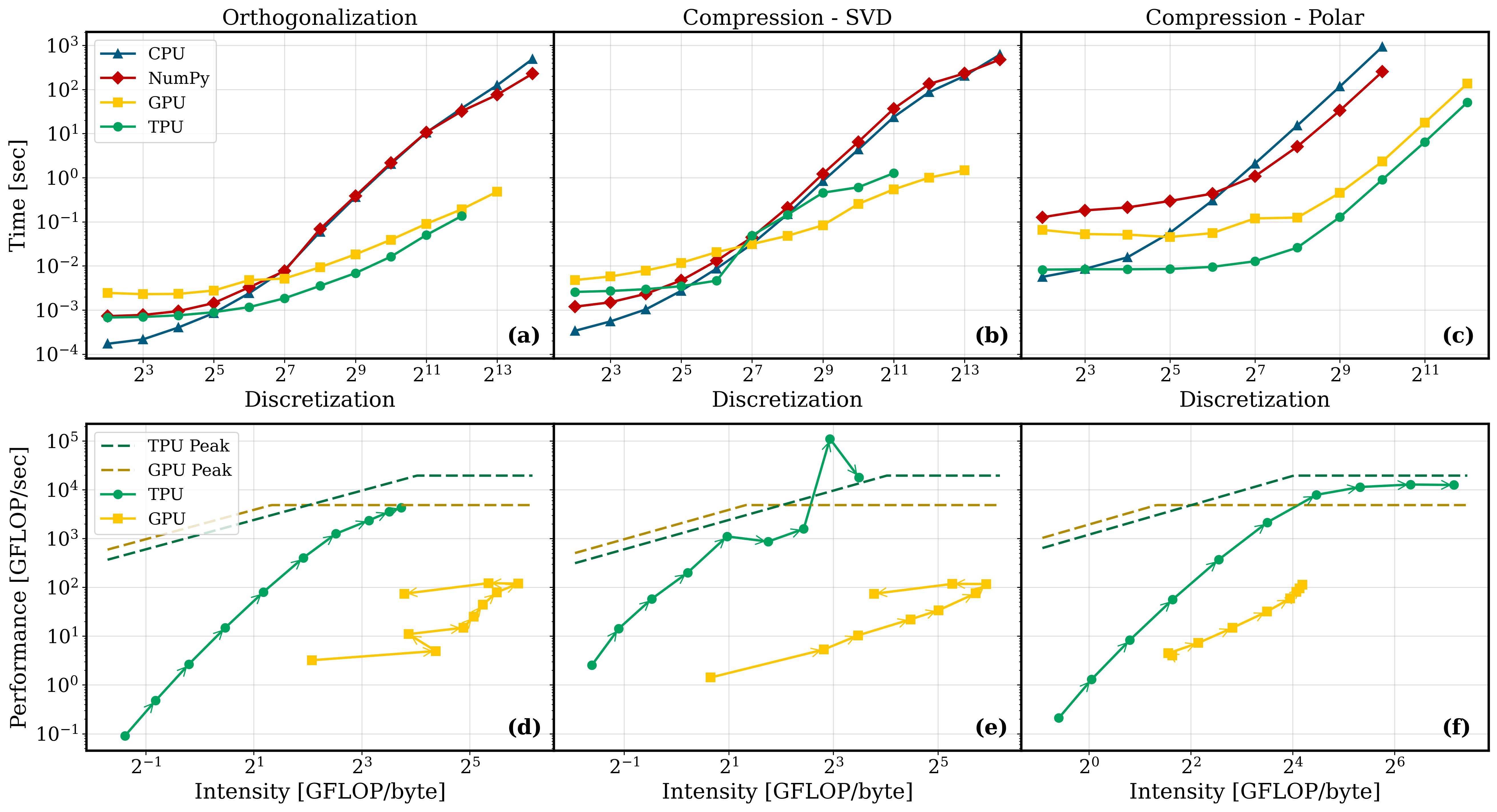}
    \caption{Serial TT operations. Columns show orthogonalization (a,d), SVD-based compression (b,e), and polar-based compression (c,f). The first row shows runtime benchmarks, the second the GPU and TPU Roofline models.}
    \label{fig:nonparallelizableTTOperations}
\end{figure*}

The figures in this section are structured in the following way:
The first row of both Fig.~\ref{fig:parallelizableTTOperations} and Fig.~\ref{fig:nonparallelizableTTOperations}
shows the average execution time as a function of the number of discretization points of the matrix represented by the underlying TT;
The second row depicts the corresponding Roofline model for both the GPU and TPU \cite{williams09}, illustrating the achieved performance $P$ in GFLOP/sec as a function of the arithmetic intensity I, measured in GFLOP/byte.
The arithmetic intensity can be obtained through 
\begin{align}
\label{eq:algorithmic_intensity}
    I = \frac{P}{B},
\end{align}
by additionally measuring the achieved memory bandwidth $B$ in byte/sec.
For visual clarity, the datapoints belonging to a given device (GPU or TPU) are connected by arrows, indicating the progression from one discretization level to the next (i.e.\ $2^2 \rightarrow 2^3 \rightarrow 2^4 \rightarrow \dots$).


The theoretical peak performance limits $\hat{P}$ of both accelerators are indicated in the figures, providing a reference to assess whether each operation is memory-bound or compute-bound.
According to the Roofline model \cite{williams09}, the achievable performance as a function of the arithmetic intensity $I$ (in GFLOP/byte) is given by
\begin{align*}
  \hat{P}(I) 
    &= \min\!\left( P_{\mathrm{peak}},\; I \cdot B_{\mathrm{peak}} \right),
\end{align*}
where $P_{\mathrm{peak}}$ denotes the compute peak (GFLOP/sec) and $B_{\mathrm{peak}}$ the memory bandwidth (byte/sec), see Appendix \ref{app:hardware} for specific values.
The \emph{ridge point} of the Roofline, obtained according to equation \eqref{eq:algorithmic_intensity} as
$I_{\mathrm{ridge}} = P_{\mathrm{peak}}/B_{\mathrm{peak}}$,
separates memory-bound operations (left of the ridge) from compute-bound operations (right of the ridge),
highlighting the dominant performance bottleneck for each TT operation.

\subsection{Parallelizable Operations}
\label{subsec:discussion_fundamentaloperations_parallel}

The first row of Fig.~\ref{fig:parallelizableTTOperations} compares the four compute modes in terms of mean iteration times for three operations on a TT: (a) addition, (b) multiplication between two TTs, and (c) multiplication between a TT and a TTO. These operations are inherently core-to-core operations and therefore highly parallelizable, suggesting the potential for substantial speed-ups on GPU architectures. Indeed, both accelerators exhibit performance improvements of several orders of magnitude over the two CPU modes, which scale exponentially with discretization.

At lower discretization levels, the TPU implementation benefits from relatively low overhead, resulting in a notable speed advantage. However, as the discretization increases, the GPU demonstrates superior scalability, eventually surpassing the TPU in all three operations at the upper end of the probed discretization range.

The second row of Fig.~\ref{fig:parallelizableTTOperations}, which presents the corresponding Roofline model, reveals that TPU performance is strongly memory-bound. As discretization increases, the achieved arithmetic intensity remains nearly constant, while the achieved performance grows significantly. This nearly vertical trend indicates that performance scales almost linearly with memory traffic, highlighting that the TPU is not only limited by memory bandwidth rather than computational throughput but also wont leave this regime for higher discretizations.

For the GPU, the Roofline model reveals an interesting behavior for the Addition operation shown in subplot (d). Both arithmetic intensity and performance initially increase with growing discretization. However, after a discretization of 
$2^7$,
a sudden drop in intensity is observed. This can likely be attributed to the L2 cache: for smaller matrices, the cache is large enough to hold all memory associated with the computation, including the input matrices, the output, and any intermediate data. Since Addition is a lightweight operation that generates relatively few additional intermediate values, the entire computation can fit within L2 for smaller discretizations. Once the matrix exceeds the L2 capacity, additional memory accesses must go to slower global memory, resulting in a decrease in arithmetic intensity.
After the drop in Intensity, the Performance follows the possible peak performance almost perfectly, suggesting that for the given resources the addition operation optimally uses the available GPU resources.
For the remaining subplots (e) and (f), the roofline model is in the compute-bound regime while its trend for growing discretizations comes close to the maximal possible performance.

\subsection{Serial Operations}
\label{subsec:discussion_fundamentaloperations_serial}

The first row of Fig.~\ref{fig:nonparallelizableTTOperations} shows the execution time of the serial operations on a TT, i.e. (a) Orthogonalization, (b) Compression via SVD and (c) Compression via polar decomposition.
The CPU-based implementations are mostly on par with each other, while the only difference occurs for smaller discretization, where the JAX-based approach has lower overhead.
Additionally, the NumPy version performs better in case of the polar-based Compression for discretizations above $2^6$.
The accelerators show a clear trend, where the TPU outperforms the GPU for the full regime in the case of Orthogonalization and Polar based Compression. However, the plot seems to indicate tat the GPU implementation for the Orthogonalization has improved growth behavior in comparison to the TPU.
A noteworthy exception is the SVD-based compression in Fig.~\ref{fig:nonparallelizableTTOperations} (b).
Here the TPU speed breaks down around $2^7$ discretization points.
This happens because the SVD on TPU working on complex64 accuracy does not converge. This can be seen in Appendix \ref{app:accuracy} were the error plots show a failure of the TPUs SVD procedure in this case.
During the evaluation similar problems occured when testing other methods like the QR decomposition with lower accuracy arithmetic like FP32.
Due to this non-convergence, the maximal number of iterations to obtain an SVD solution is reached, resulting in the sudden increase in execution time.

Similar behavior can be observed in subplot (e) of the second row in Fig.~\ref{fig:nonparallelizableTTOperations}. 
Here, the TPU performance fluctuates for discretizations beyond $2^7$ and even exceeds the theoretical peak performance. 
This is likely caused by measurement inaccuracies of the JAX Profiler under this regime.  

The trends for the remaining subplots, (d) and (f), corresponding to Orthogonalization and Polar compression respectively, 
are similar: they start in the memory-bound regime and gradually move toward higher arithmetic intensities and achieved performance as the discretization increases. 
While the compression operation approaches the boundary defined by the Roofline ridge point, the Polar-based compression algorithm transitions into the compute-bound regime, reaching values close to the theoretical upper limit achievable on the TPU.  

GPU performance in the mostly serial regime is more variable, which is expected since GPUs are less efficient for predominantly serial workloads. 
However, subroutines within these operations—such as QR or SVD decompositions—may still exhibit partial parallelism and benefit from GPU acceleration. 
For subplots (d) and (e), the GPU Roofline model shows initial growth in intensity and performance with increasing discretization until a saturation point is reached. Beyond this point, the arithmetic intensity decreases significantly while performance drops only slightly.  
This suggests an implicit upper bound on the achievable performance for these mostly serial routines. 
The true peak performance in this regime lies well below the theoretical maximum, reflecting the limited parallelism. 
The observed deterioration in intensity can be attributed to a combination of increased memory traffic (e.g., due to padding for larger matrices) and low GPU occupancy, which causes the Streaming Multiprocessors (SMs) to remain idle while waiting for the few active threads to execute.  

The last subplot (f) exhibits similar initial behavior for the GPU but has not yet reached the point of intensity deterioration. 
Instead, the Roofline trajectory appears to converge near the boundary between the memory-bound and compute-bound regimes.  

Overall, the data does not indicate a clear superiority of one accelerator over the other. 
The choice of hardware depends on the specific workload and the balance between serial and parallelizable fundamental TT operations in a given application.
However, both accelerators are orders of magnitudes faster than reference implementations performed on CPU.


\begin{figure}
    \centering
    \includegraphics[width=\linewidth]{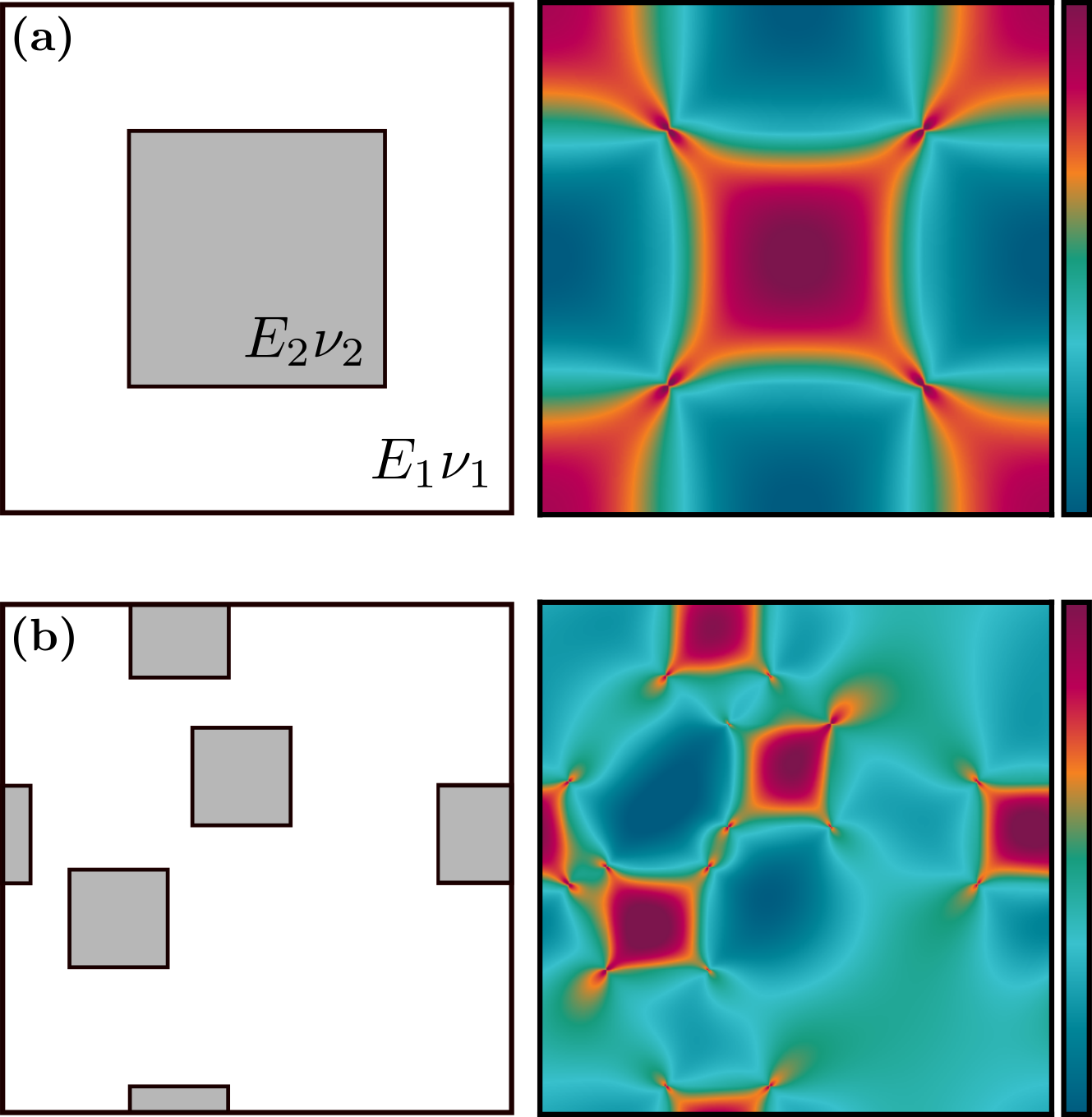}
    \caption{Left: geometries used in the experiments. Right: local stress field $\sigma_{xy}$ under tensile loading obtained with the quantum-inspired homogenization algorithm on TPU. (a) box and (b) multibox geometries. Material parameters are $E_1 = 29/3\,\si{\giga\pascal}$, $E_2 = 4/3\,\si{\giga\pascal}$, with $\nu_1 = \nu_2 = 1/3$ for the grey and white phases, respectively.}
    \label{fig:geometries}
\end{figure}

\section{Accelerated Quantum \\Inspired Homogenization}
\label{sec:acceleratedQuantumInspiredHomogenization}
In this section, we extend the insights obtained from the preceding analysis to a realistic application scenario involving the SFFT-based homogenization algorithm \cite{hauck26}. We benchmark the JAX list-based implementation on CPU, GPU, and TPU architectures, using a GPU-accelerated implementation of the classical FFT-based homogenization algorithm as a baseline reference. The latter is implemented using the CuPy package \cite{cupy17} and builds on cuFFT \cite{cufft}, NVIDIA’s CUDA-based FFT library.
As in the original formulation of the SFFT-based scheme, we employ an adaptive-rank procedure.

Additionally, we include warm-up times in the final benchmarking procedure. This is necessary because, within the adaptive-rank setting, the final tensor ranks are not known a~priori. Consequently, the shapes of intermediate tensors cannot be predetermined. These shapes emerge dynamically during the execution of the algorithm and will trigger new additional JIT compilations.

For the application of the homogenization algorithm, we restrict ourselves to the two-dimensional box and the multibox geometry shown in Fig.~\ref{fig:geometries}(a) and (b), respectively. We set the compression accuracy to $\epsilon = 10^{-5}$ for all runs; however, the operators themselves are computed with a higher accuracy of $\epsilon = 10^{-10}$ to ensure numerical robustness.




A central step within the original SFFT-based homogenization algorithm is the repeated use of SVD-based compression after each step in the algorithm to recompress the underlying TT. In high-discretization regimes and under reduced-precision arithmetic, we observe that the SVD procedure becomes numerically unstable when performed on the TPU and may fail to converge within the prescribed iteration budget.
To address this issue, we replace the standard SVD-based compression with the polar decomposition-based alternative, but solely for the TPU implementation. The polar decomposition exhibits substantially improved numerical stability under low-precision arithmetic, as it avoids the iterative singular value computation and instead relies on a more stable matrix factorization with better-conditioned convergence behavior in practice, see Appendix~\ref{app:accuracy}.

\subsection{Coarse-Graining Approach}
\label{sec:coarse_graining}
Directly recompiling a jitted function for each rank that may arise during the simulation is inefficient, as the accumulated warm-up times would outweigh any potential speed-up. To mitigate this issue, we introduce a coarse-graining strategy for the TT-ranks.

Instead of allowing each core rank $r_c$ to assume any positive integer value, we restrict it to the form
$r_c = n\, r_0$ ($n, r_0 \in \mathbb{N}^+$), where $r_0$ is a prescribed minimal base rank and $n$ denotes its multiplicity. These parameters are chosen to reduce the number of distinct tensor shapes encountered during execution, thereby minimizing recompilation events while still retaining the benefits of rank adaptivity. Selecting $r_0$ equal to or larger than the naturally emerging maximal rank $r$ would eliminate adaptivity entirely and effectively yield a fixed-rank method.

Coarse-graining is introduced through a lightweight modification of the compression algorithm. Instead of adapting the rank dimension iteratively during compression, we first set all singular values above the truncation threshold to zero and record the corresponding threshold index. After collecting these truncation indices across all cores, a separate coarse-graining function applies the truncation according to the aggregated parameters. This separation of responsibilities ensures that tensor shapes are preserved within the jitted compression routine.
Importantly, no further modifications are required for other TT operations. Addition, multiplication, and related algebraic operations naturally produce new TT representations whose intermediate ranks follow from the algebraic structure of the involved cores.

For our experiments, we set the minimal base rank to $r_0 = 16$. This value could be further refined by allowing $r_0$ to depend on the discretization rather than treating it as a constant.

\begin{figure}
    \centering
    \includegraphics[width=\linewidth]{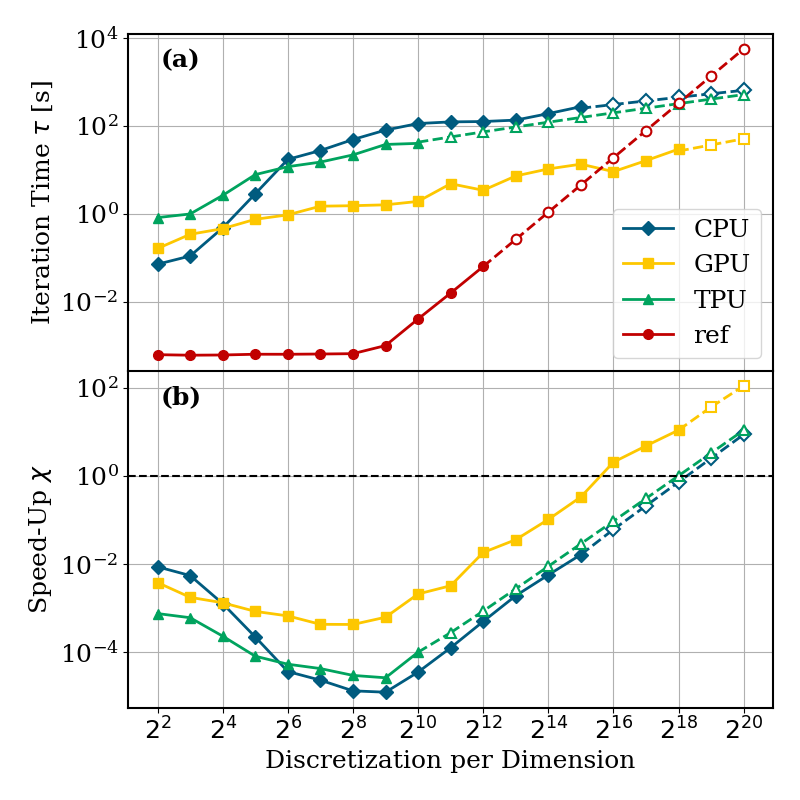}
\caption{Performance of the SFFT-based homogenization algorithm on the box geometry: (a) average iteration times $\tau$; (b) achieved speed-up $\chi$. The colored-dashed lines indicates extrapolations while the dashed-black line corresponds to the break-even point.}
\label{fig:speed_box}
\end{figure}

\begin{figure}
    \centering
    \includegraphics[width=\linewidth]{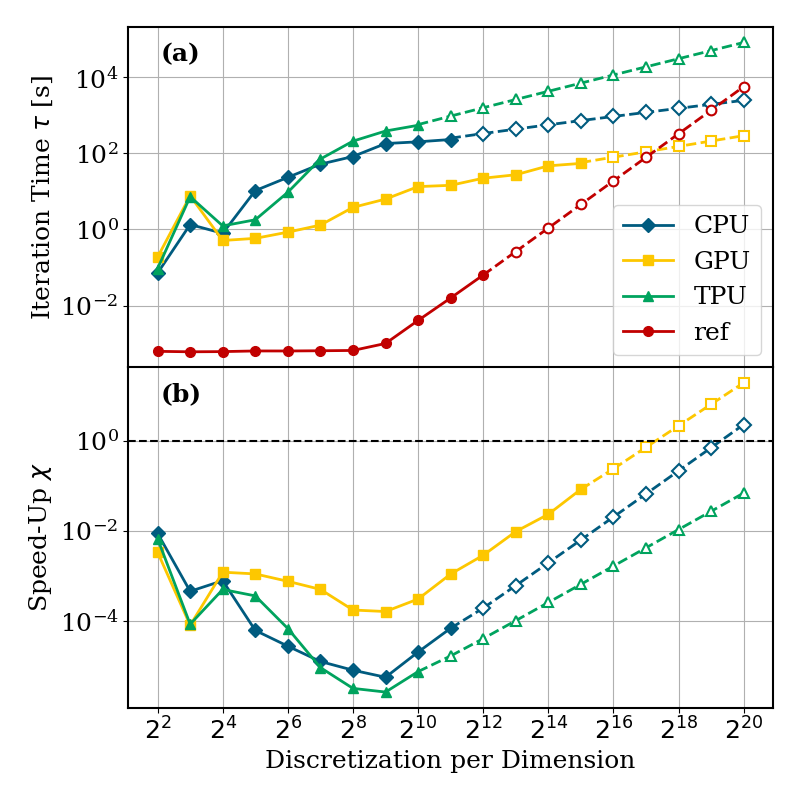}
    \caption{Performance of the SFFT-based homogenization algorithm on the multibox geometry: (a) average iteration times $\tau$; (b) achieved speed-up. The colored-dashed lines indicates extrapolations while the dashed-black line corresponds to the break-even point.
    }
    \label{fig:speed_multibox}
\end{figure}

\subsection{Performance Analysis}
Running Algorithm~\ref{algo:sfft} on different compute devices requires the same number of iterations to reach the tolerance $\delta_{\mathrm{tol}} = 10^{-4}$. Consequently, the key performance metric is the average time per iteration, $\tau$, as well as the achieved speed-up
$\chi = \tau_{\mathrm{cuFFT}} / \tau_{\mathrm{QTT}}$, which measures the performance of the SFFT-based homogenization algorithm on a given platform relative to the cuFFT-based reference.

The results for the box geometry are shown in Fig.~\ref{fig:speed_box} (a). 
The cuFFT-based implementation shows superior performance for small discretizations but exhibits FFT-characteristic scaling beyond $2^9$ points per dimension (ppd). Due to memory limitations, it cannot exceed $2^{12}$ ppd, see section \ref{subsec:stability_scaling}.
In contrast, the GPU-based quantum-inspired solver scales up to $2^{18}$ ppd (approximately $70$ billion grid points), while the CPU and TPU versions reach $2^{14}$ ppd and $2^{10}$ ppd, respectively, reflecting their more limited memory capacity.
A crossover with the cuFFT baseline is observed for the GPU variant under extrapolation and is expected to remain asymptotically faster at even higher discretizations. This behaviour is also observed for the CPU and TPU; however, the TPU implementation shows slightly better overall performance than the CPU implementation but remains more than one order of magnitude slower than the GPU version.

Simulating the more complex multibox geometry, see Fig.~\ref{fig:speed_multibox}(a), shows the same overall trend for the cuFFT baseline as for the box case due to its geometry-agnostic formulation. The quantum-inspired solver also exhibits similar scaling behavior as in the box case; however, its memory scaling is slightly worse, limiting the GPU implementation to $2^{14}$ ppd (approximately $270$ million pixels), as discussed in the following section. Due to increased memory demands, the TPU reaches only $2^{9}$ ppd, while the CPU achieves up to $2^{10}$ ppd. The TPU variant shows slightly worse performance and asymptotic scaling than the CPU implementation, which can partially be attributed to the missing leading data point at $2^{10}$ ppd.

For all platforms, the speed-up curves for both geometries (Fig.~\ref{fig:speed_box}(b) and \ref{fig:speed_multibox}(b)) show a positive slope once FFT scaling dominates the cuFFT based implementation. This suggests that a crossover point will eventually be reached where each platform outperforms cuFFT implementation, even under conservative scaling assumptions.

It should be noted that the reference implementation used in the above simulations is based on highly optimized FFT libraries. In contrast, the code developed for the SFFT-based homogenization approach is considerably less optimized.
Therefore, the reported execution times are not fully representative of the method’s ultimate performance. The main takeaway is thus the observed scaling behavior rather than the absolute timing values.

\subsection{Memory and Stability Analysis}
\label{subsec:stability_scaling}

The principal advantage of the TT representation lies in its compressed storage complexity. For geometries with approximately separable structure, the TT ranks remain moderate even under increasing discretization, enabling substantially larger resolutions than dense FFT-based approaches.

\begin{figure}
    \centering
    \includegraphics[width=\linewidth]{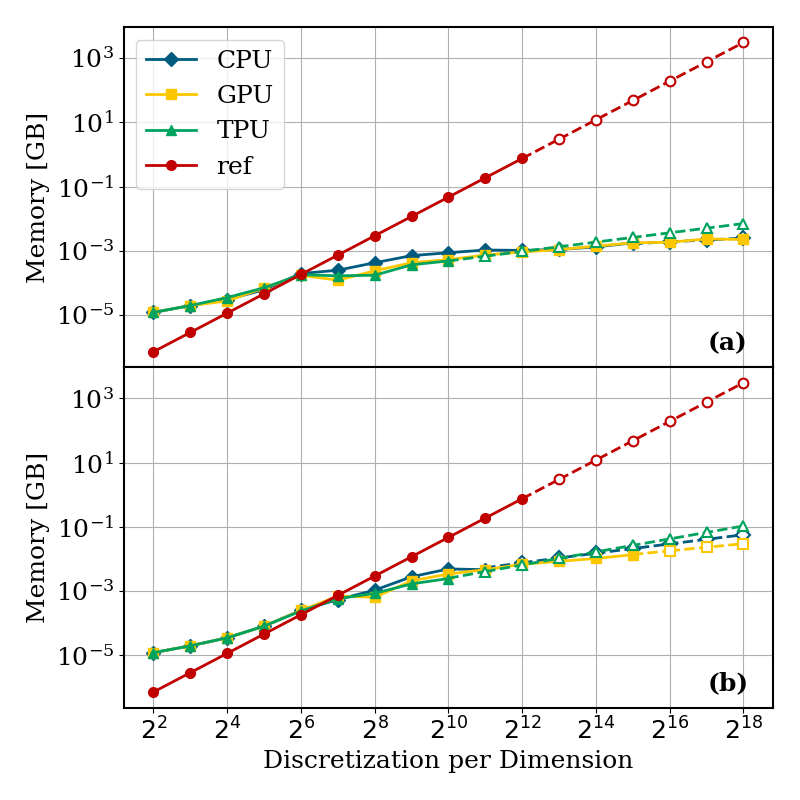}
    \caption{Memory needed to represent the final strain field for (a) the box and (b) the multibox geometry. The dashed lines indicates extrapolations.}
    \label{fig:memory}
    
\end{figure}

Figure~\ref{fig:memory} reports the memory consumption for the strain calculated through the SFFT-based homogenization algorithm on CPU, GPU, and TPU architectures, as well as the cuFFT-based reference implementation, for varying discretizations. While the cuFFT formulation exhibits growth characteristic of the underlying dense representation, the TT formulations show significantly reduced and only marginal memory growth beyond $2^{6}$ ppd.
Notably, the box geometry exhibits almost one order of magnitude lower memory consumption than the multibox geometry at equal discretization.

Nevertheless, the achievable problem size is ultimately limited by rank growth. For the geometries considered in this work, the ranks remain stable across higher resolutions. In contrast, strongly curved or non-axis-aligned geometries exhibit non-saturating rank growth, leading to earlier memory saturation and OOM events.
A further source of memory overhead arises from intermediate tensor contractions, which may temporarily exceed the final compressed tensor size. Consequently, the memory footprint is governed not only by the final TT representation but also by transient contraction complexity during execution. Additionally, platform-dependent JAX/XLA compilation can produce different computation graphs, including variations in operator fusion and memory layout, which can significantly affect peak memory usage and potentially contributes to the earlier OOM events observed on CPU and TPU.

The most important figure of merit for the Homogenization algorithm is the accuracy of the homogenized stiffness tensor. As an error metric, we consider the relative Frobenius norm error
\begin{align*}
    \overline{\varepsilon} =\frac{||C^{TT}-C^{ref}||_F}{||C^{ref}||_F}
\end{align*}
where $\mathbf{C}^{\mathrm{TT}}$ denotes the homogenized stiffness tensor obtained from the compressed TT formulation and $\mathbf{C}^{\mathrm{ref}}$ corresponds to the dense cuFFT based reference.

\begin{figure}
    \centering
    \includegraphics[width=\linewidth]{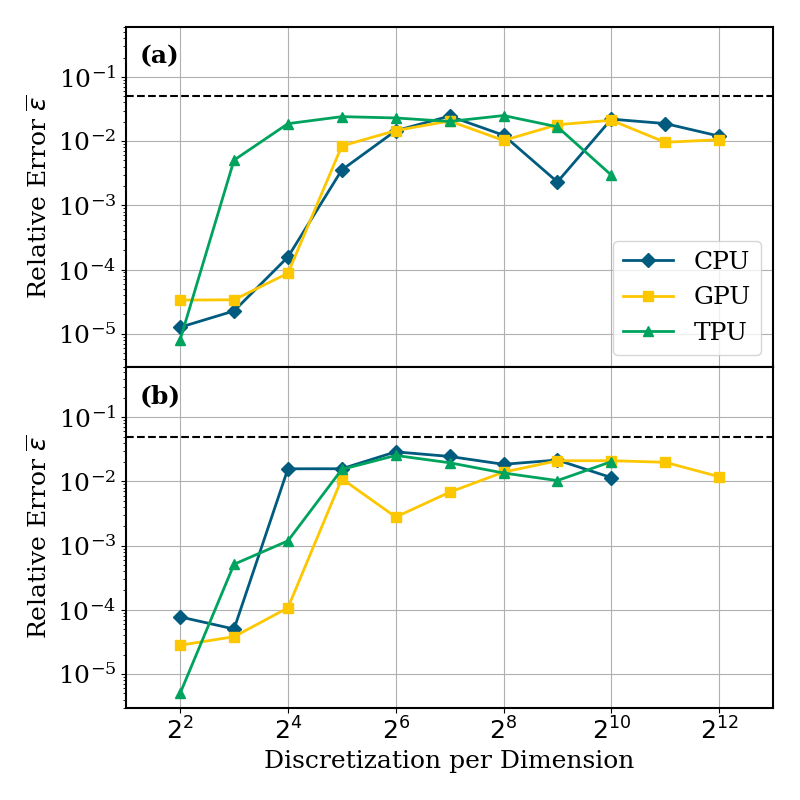}
    \caption{Relative error $\overline{\varepsilon}$ in the final stiffness tensor for (a) the box geometry and (b) the multibox geometry. The dashed line indicates the industrial 5\% threshold.}
    \label{fig:error}
\end{figure}

Figure~\ref{fig:error} shows the relative error as a function of discretization size for both geometries. The accuracy is controlled by the cutoff parameter $\epsilon$, chosen such that the TT-based approach maintains a relative error below 5\%, corresponding to a practically acceptable accuracy level for effective material properties. Decreasing $\epsilon$ further improves accuracy at the cost of increased memory consumption. The reported data-points lie within the regime where dense reference solutions remain computationally feasible.


\subsection{Data-driven approaches for computational homogenization}

We conclude this paper by briefly place the SFFT-based homogenization framework within the broader context of data-driven multiscale methods. Recent advances in machine learning, in particular neural operators such as Fourier Neural Operators, have enabled efficient learning of mappings from microstructures to effective macroscopic responses \cite{azizzadenesheli24, li21, kovachki23}. While these approaches offer fast inference once trained, they typically require large-scale high-fidelity datasets, shifting much of the computational cost to an offline training phase \cite{choubey25}.

In this setting, the proposed SFFT-based method can serve as a complementary tool by efficiently generating accurate reference solutions for diverse microstructures, which can then be used to train or validate data-driven surrogate models. Unlike purely learned approaches, it enforces the governing equations directly at each evaluation, ensuring physical consistency without the need for training data. This combination suggests a hybrid workflow in which physics-based quantum-inspired solvers and data-driven models jointly enable scalable and reliable multiscale analysis.

\section{Conclusion}
\label{sec:conclusion}
In this work, we presented two major contributions advancing the practical use of Tensor Train methods on modern hardware accelerators and demonstrating their applicability to large-scale, quantum-inspired computational homogenization.

Our first contribution is a systematic, cross-platform benchmarking study of fundamental Tensor Train operations on CPUs, GPUs, and TPUs using the JAX and NumPy framework. 
The results show that GPUs outperform TPUs primarily for highly discretized problems that are inherently parallelizable. 
For smaller problem sizes, as well as for operations with a more serial computational structure, TPUs consistently exhibit superior performance across the tested regime.
While GPUs are predominantly compute-bound throughout the benchmark, TPU performance varies: it is strictly memory-bound for highly parallel operations, and transitions from memory-bound at small problem sizes to compute-bound at larger discretizations for more serial tasks. 
Despite being a less mature hardware platform, TPUs delivered notable performance in our tests. 
We did, however, observe numerical instabilities on TPUs for SVD operations under \texttt{complex64}, though these limitations are expected to diminish as the TPU ecosystem continues to mature.

Our second contribution applies these insights to the implementation of the quantum-inspired homogenization algorithm under realistic high-memory conditions. Running a fully rank-adaptive algorithm in JAX required several modifications, including a coarse-graining strategy to mitigate the substantial warm-up costs associated with JIT compilation. Even under these constraints, GPUs achieve substantial and, in practical regimes, exponentially increasing speed-ups over the cuFFT-based reference implementation.
The TPU implementation shows a similar performance to the JAX-based CPU implementation, which can mainly be attributed to the need for a more stable but slower polar-based compression algorithm instead of the favorable SVD-based approach.
A remaining practical limitation is the JAX warm-up time: each admissible coarse-grained rank triggers additional compilations during the early stages of the algorithm. 
However, upcoming JAX features enabling persistent caching of compiled functions, together with the possibility of treating JIT compilation as a preprocessing step, are expected to substantially reduce this overhead.  
At the time of writing, these features remain experimental. 
Further speed-ups are anticipated once the SVD-based routines in high-discretization settings,
become numerically robust on TPUs.

Although this work does not yet constitute fully industrial-scale homogenization for arbitrary microstructures, the results indicate a viable pathway toward such applications for classes of approximately low-rank geometries. In particular, pixelized microstructures arising from layered composites, additive manufacturing toolpaths, lattice materials, or topology-optimized structures often exhibit significant spatial correlations that are compatible with the TT format.



Future work may extend these ideas in several promising directions. Higher-order tensor network architectures, such as Tree Tensor Networks or Projected Entangled Pair States, may further reduce memory requirements and improve the representational capacity of the proposed SFFT-based homogenization framework, particularly in cases where current low-rank structures become restrictive.
In addition, the development of a batched and hardware-accelerated quantum-inspired homogenization pipeline would enable efficient evaluation over large ensembles of microstructures.
These directions point toward scalable hybrid computational frameworks that integrate tensor network compression with hardware-accelerated and data-driven homogenization strategies for multiscale material modeling.

\section{Acknowledgment}
The authors thank Dr. Mazen Ali for many valuable and insightful discussions. 
This research was supported in part through the use of Cloud TPUs provided by Google’s 
TPU Research Cloud (TRC) program.
Further, we thank BMWK for the financial support provided under the EniQmA project.
The authors gratefully
acknowledge funding from the German National High Performance
Computing (NHR) association for the Center NHR South-West.
To enhance readability this manuscript was refined with the assistance of FhGenie. 

\appendix
\section{Extracting the Stiffness\\ Tensor}
\label{app:extractingStiffness}

Working with full tensor representations can be cumbersome both in terms of clarity and computational efficiency.  
We therefore adopt Voigt notation, representing stress and strain tensors as 3-dimensional vectors in 2D space.
The stress and strain tensors are written in vector form as
\[
\boldsymbol{\sigma} =
\begin{bmatrix}
\sigma_{11} & \sigma_{22} & \sigma_{12}
\end{bmatrix}^{\mathsf{T}},
\qquad
\boldsymbol{\varepsilon} =
\begin{bmatrix}
\varepsilon_{11} & \varepsilon_{22} & 2\varepsilon_{12}
\end{bmatrix}^{\mathsf{T}}.
\]
This convention can be extended to the fourth-order stiffness tensor, which in Voigt representation is mapped to a $3 \times 3$ matrix.
It should be noted that the factor of 2 for the off-diagonal strain component is included to ensure that the free energy of the system
\begin{align*}
    F = \frac{1}{2} \sum_{i=1}^3 \sum_{j=1}^3 \sigma_{ij} \epsilon_{ij}
\end{align*}
is correctly represented \cite{sadd09, schwarz13}. This is purely a convention and could, in principle, be defined differently.  

Using Voigt notation, the macroscopic strain-stress relationship \eqref{eq:macro_ss} can be written as a simple matrix-vector product, greatly simplifying computations.  
By setting the mean strain vector $\mathbf{E}$ such that only its $k$-th component is unity, we further obtain
\begin{align*}
\sigma^\mathrm{eff}_i = \mathbf{C}^\mathrm{eff}_{ij} \delta_{jk} = \mathbf{C}^\mathrm{eff}_{ik},
\end{align*}
allowing the $k$-th column of $\mathbf{C}^\mathrm{eff}$ to be read directly from the macroscopic stress vector.  
Repeating this procedure for all components of $\mathbf{E}$ constructs the full effective stiffness matrix systematically. 
To determine the macroscopic stress, 
the mean strains are provided as inputs to Algorithm~\ref{algo:fft}.
Since the output is formed by the local stresses, we can averaging over the RVE to obtain the needed macroscopic stresses 
\begin{align*}
    \sigma^\mathrm{eff} =
    \frac{1}{|\mathcal{V}|} \int_\mathcal{V} \sigma_m(\mathbf{x}) \, \mathrm{d}\mathbf{x}
    =
    \frac{1}{|\mathcal{V}|} \int_\mathcal{V} \mathbf{C}(\mathbf{x}) \, \varepsilon_m(\mathbf{x}) \, \mathrm{d}\mathbf{x}.
\end{align*}




\section{SVD- vs Polar-based\\ Compression Benchmark}
\label{app:SVD_vs_Polar}

\begin{figure}
    \centering
    \includegraphics[width=0.95\linewidth]{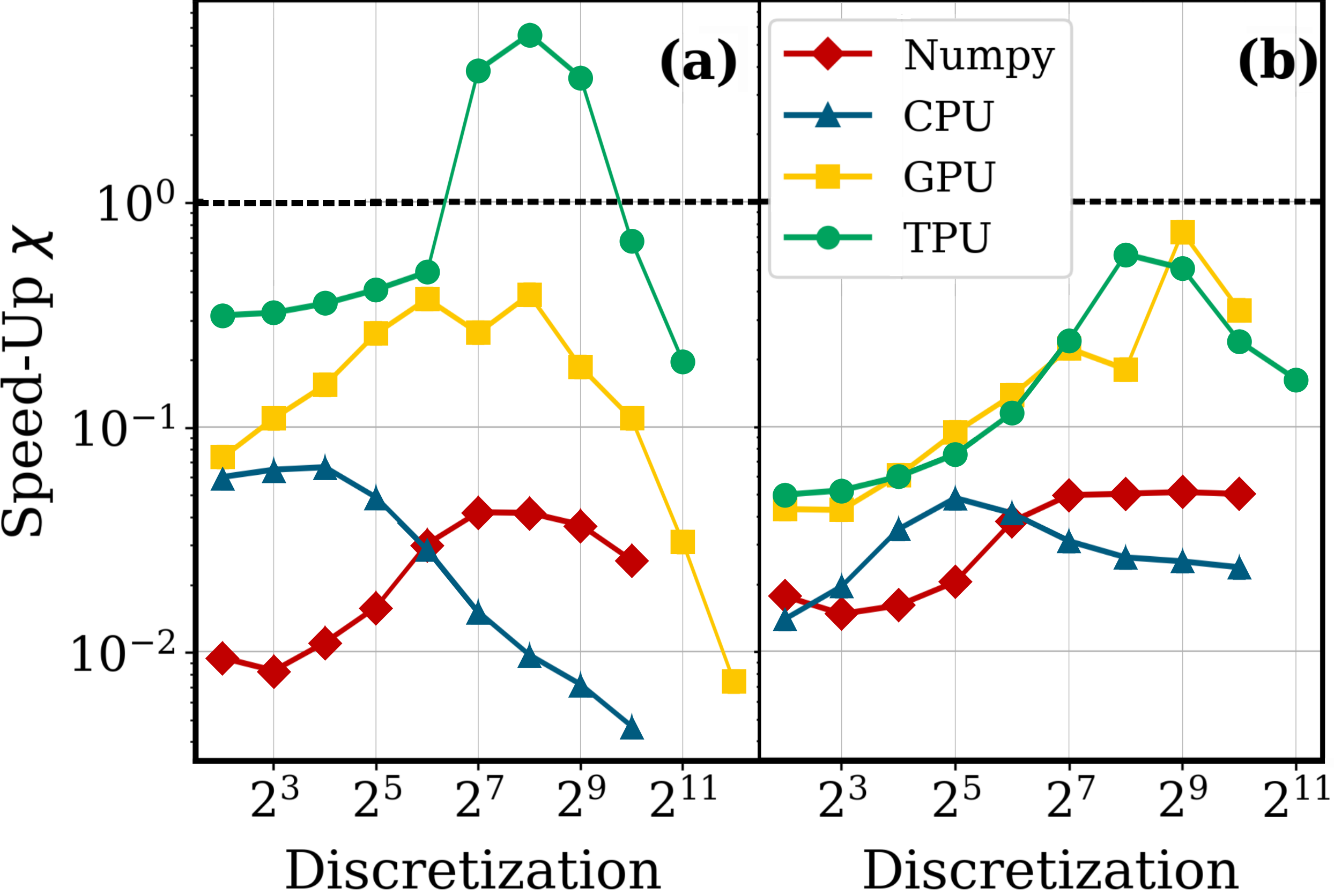}
    \caption{Speed-Up Factors between the Polar- and SVD-based Compression methods for both (a) the list-based implementation and (b) the batched implementation.
    The dashed-black line corresponds to the break-even point.}
    \label{fig:SVDvsPolar}
\end{figure}

As suggested in the main text, the SVD-based Compression should be significantly faster than the Polar-based approach. However, only as long as the underlying matrix is well behaved. 
The results obtained for the relative speed-up
$\chi = \frac{\tau_\mathrm{Polar}}{\tau_\mathrm{SVD}}$
defined through the mean single iteration time for the SVD- and Polar-based compression, $\tau_\mathrm{SVD}$ and $\tau_\mathrm{Polar}$ respectively,
for both the list-based and batched format are shown in Fig. \ref{fig:SVDvsPolar}.
It can be seen that only the TPU implementation in the List format can cross the break-even point at $\chi=1$.
Thus, only for this specific case we can actually get a faster implementation if one would choose the Polar-based variant.
This however, only occurs at and after a discretization of $2^7$, which is the same discretization at which the SVD-based algorithm starts to fail anyway, see Appendix \ref{app:accuracy}.
Thus, we can conclude that the SVD-based approach is preferable over the Polar-based algorithm as long as it is applicable.
We can further see that the batched version has smaller variations in its speed-up factor and seems to even become stable for the JAX and NumPy based CPU implementations.
Hence, our initial hypothesis on the SVD- and Polar-based conundrum are correct for all checked compute platforms.

\begin{figure*}[tbph]
    \centering
    \includegraphics[width=\textwidth]{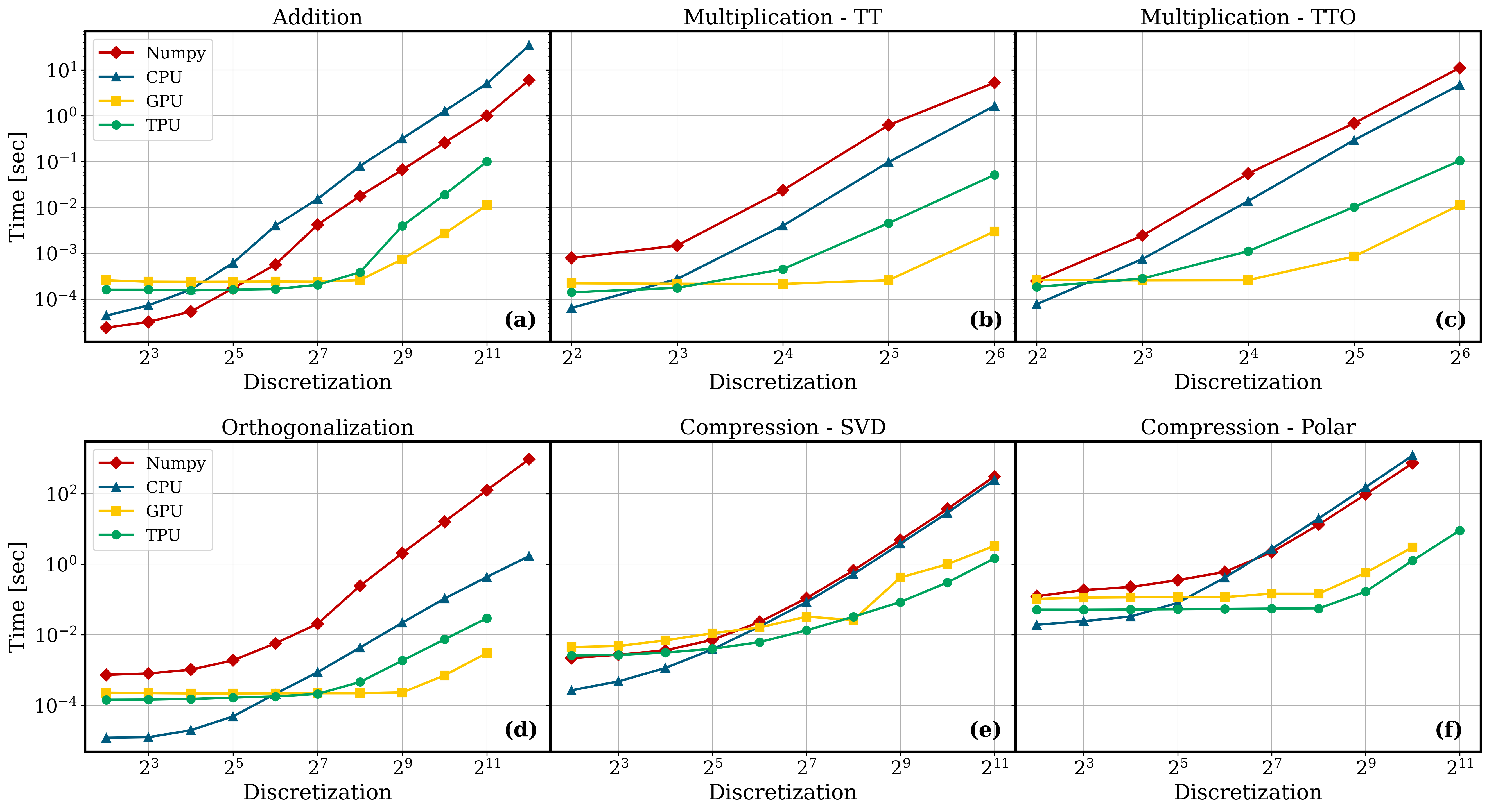}
    \caption{Time for the JAX batched variant as alternative implementation to the JAX list-based TT version.}
    \label{fig:AbsoluteTimingBatched}
\end{figure*}

\section{Alternative Tensor Format}
\label{app:jax_tensor_Homogenization}

In this section we want to give additional benchmarking results obtained for the batched format and further compare them with the used list-based format in terms of achieved speed-up.
For this, we define the speed-up $\chi = \frac{\tau_\mathrm{batch}}{\tau_\mathrm{list}}$,
with the mean run times $\tau_\mathrm{list}$ and $\tau_\mathrm{batch}$ for the list-based and batched implementations, respectively.

The results for the absolute time obtained per benchmarked operation are shown in Fig.~\ref{fig:AbsoluteTimingBatched}.
It can be seen that the general trend for the different operations and compute platforms follow a similar trend than the ones obtained for the list-based implementations shown in Fig. \ref{fig:nonparallelizableTTOperations} and \ref{fig:parallelizableTTOperations}. 
Concerning the JAX-CPU and NumPy versions of the shown operations, it becomes clear that the JAX-implementation can be significantly faster in comparison to the list-based implementation. 
This shows the improvement that JAX can have on the optimization of the underlying code in the case of ever growing matrices.
We can see further that the TPU seems to reach its maximal potential -- indicated by a near zero gradient -- for smaller discretizations as for the list-based approach. 
This is not seen for the GPU version, suggesting that in the batched format GPUs are superior.

\section{Accuracy Results}
\label{app:accuracy}

Due to clarity we want to share the results obtained for the numerical accuracy obtained for the list-based implementation for the different probed operations.
The figure of interest is the relative error defined through
\begin{align}
    \sigma_\text{err} = \frac{||\alpha_\mathrm{JAX}-\alpha_\mathrm{NumPy}||_2}{||\alpha_\mathrm{NumPy}||_2},
\end{align}
where $\alpha_\mathrm{JAX}$ is the result obtained by running either the JAX implementation on CPU, GPU or TPU.
Analogously $\alpha_\mathrm{NumPy}$ is the result obtained from the reference NumPy implementation performed on CPU.

\begin{figure*}[tbph]
    \centering
    \includegraphics[width=\textwidth]{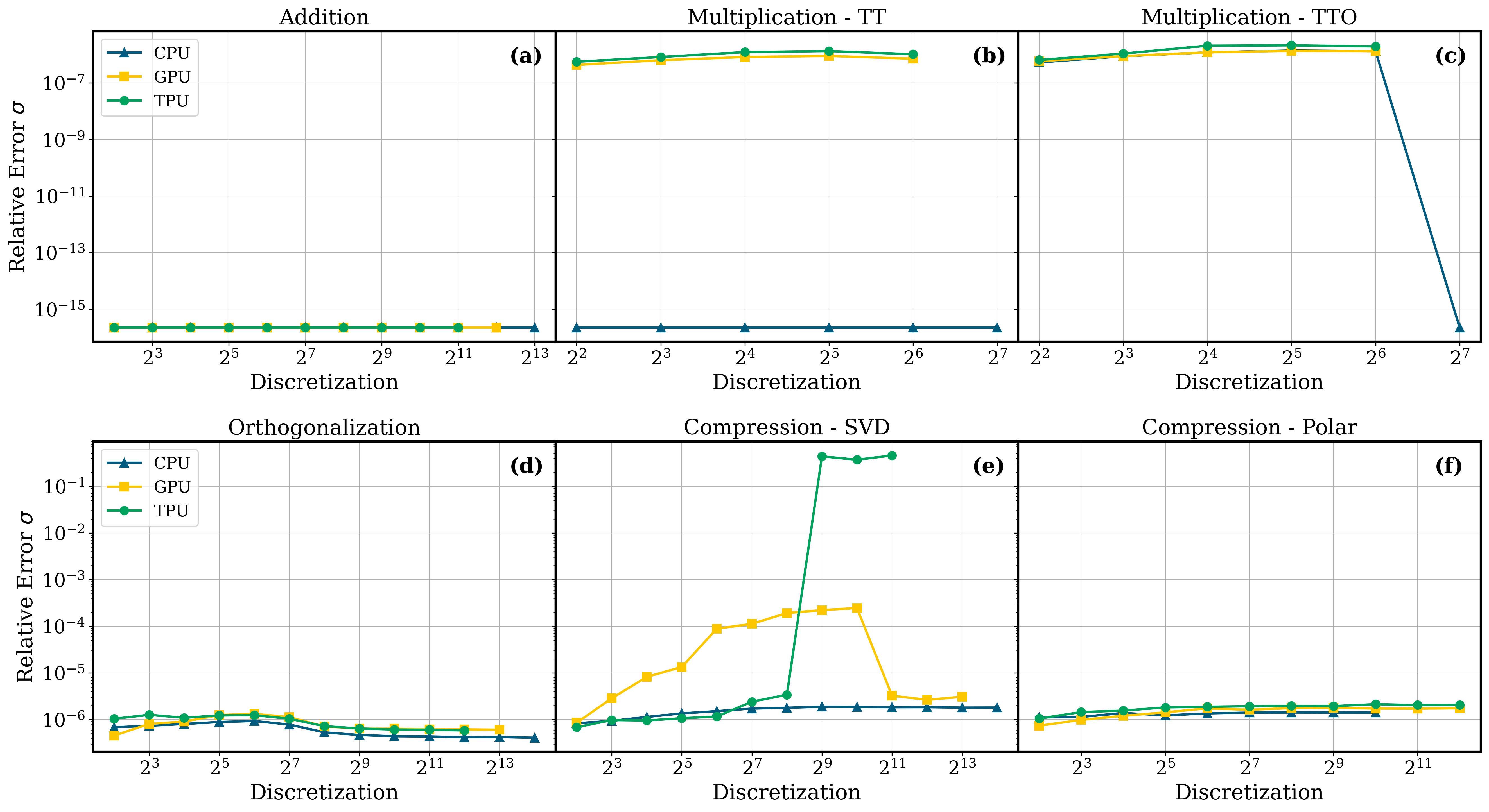}
    \caption{Relative Errors between the JAX and NumPy implementations on different Hardware.}
    \label{fig:accuracy_list}
\end{figure*}

The results of the relative error $\sigma$ for different compute platforms is shown in Fig. \ref{fig:accuracy_list}.
Most results are either around machine precission or at least below $10^{-6}$.
However, the only result that breaks with this is subplot (e) showing the SVD-based compression. 
First, the result for the GPU suggest an instability for the intermediate regime.
This may occur because the GPU pads incoming matrices to specific sizes to make optimal use of them.
However, if the padding is quite large, this may result in bad conditioning and worse performance of the SVD algorithm.
The second problem arises from the TPU version of JAX.
Here the error begins to rise around a discretization around $2^7$ and breaks down complete around $2^9$.
This suggests that the SVD algorithm didn't converge in this case and making the algorithm effectively unusable on TPU.

\section{Hardware Specifications}
\label{app:hardware}
The tables \ref{tab:cpu} - \ref{tab:tpu} summarize the hardware specifications of the compute nodes used for the experiments.

\textit{Execution setup:}
CPU and GPU experiments were run on dedicated compute nodes of the Beehive Cluster at Fraunhofer ITWM.
Unless otherwise stated, each experiment was executed on a single compute node (single CPU node or 1× NVIDIA A100 GPU). TPU experiments were conducted on a single TPU v4-8 slice provided via the Google TPU Research Cloud.

\textit{Software stack:}
All experiments were implemented in \texttt{Python 3.11} using \texttt{CUDA 12.9} and \texttt{cuFFT 12.4} backend libraries where applicable. The experiments were conducted in \texttt{JAX 0.6} using XLA compilation.

\textit{Precision and numerical settings:}
All experiments were performed in FP32 precision, with Complex64 used for complex-valued computations. GPU runs additionally employed mixed precision with TF32 enabled on Ampere architectures. Deterministic execution was not enforced, prioritizing computational throughput over bitwise reproducibility.


\begin{table}[h!]
\centering
\caption{CPU node hardware specifications.}
\label{tab:cpu}

\begin{tabular}{p{0.4\columnwidth} p{0.5\columnwidth}}
\hline
\textbf{Component} & \textbf{Specification} \\
\hline
Cluster & MPI3 \\
Compute nodes & 108 (Dell PowerEdge C6420) \\
CPU & Dual Intel Xeon Gold 6240R \\
Cores per node & 48 \\
RAM per node & 384 GB (8 GB per core) \\
Storage & 480 GB SSD \\
Network & 10 GiB Ethernet, HDR100 Infiniband \\
\hline
\end{tabular}
\end{table}


\begin{table}[h!]
\centering
\caption{NVIDIA A100 80GB PCIe system specifications \cite{nvidia21}.}
\label{tab:gpu}
\begin{tabular}{p{0.4\columnwidth} p{0.5\columnwidth}}
\hline
\textbf{Component} & \textbf{Specification} \\
\hline
Architecture & Ampere \\
GPU Memory & 80 GB HBM2e \\
Memory Bandwidth & 1{,}935 GB/sec \\
Peak Performance & 9.7 TFLOP/sec (FP64) \\
                 & 19.5 TFLOP/sec (FP32)\\
Tensor Core Performance & 19.5 TFLOP/sec (FP64) \\
                        & 156 TFLOP/sec (FP32) \\
                        & 312 TFLOP/sec (BF16 or FP32)\\
                        & 624 TFLOP/sec (int8)\\
Form Factor & PCIe dual-slot air-cooled or \\
            &single-slot liquid-cooled \\
Interconnect & PCIe Gen4 (64 GB/s) \\
TDP & 300 W \\
\hline
\end{tabular}
\end{table}

\begin{table}[h!]
\centering
\caption{TPU v4.8 system specifications \cite{chern22, jouppi23}}
\label{tab:tpu}
\begin{tabular}{p{0.4\columnwidth} p{0.5\columnwidth}}
\hline
\textbf{Component} & \textbf{Specification} \\
\hline
TPU type & v4.8 \\
Peak performance & 275 TFLOP/sec \\
&(BF16 or int8) \\
HBM2 capacity \& bandwidth & 32 GiB, 1,200 GB/s \\
Power consumption (min/avg/max) & 90/170/192 W \\
TPU-Pod size & 4,096 chips \\
Interconnect topology & 3D-Mesh \\
Peak compute per Pod & 1.1 EFLOP/sec \\
&(BF16 or int8) \\
All-Reduce bandwidth per Pod & 1.1 PB/sec \\
Shared bandwidth per Pod & 24 TB/sec \\
RAM per node & 400 GB \\
\hline
\end{tabular}
\end{table}

Since Google does not provide official peak throughput values for complex64 arithmetic on TPU v4.8 systems, we estimate performance based on the published BF16 peak throughput.
A TPU slice reaches up to
$P_{\mathrm{BF16}} = 275 \ \text{TFLOP/sec}$.
Assuming FP32 throughput is roughly half of BF16 throughput (a common precision-scaling heuristic for TPU MXUs), and using the standard cost of eight real floating-point operations per complex multiply–accumulate, we estimate
\[
P_{\mathrm{c64}} \approx \frac{P_{\mathrm{BF16}}/2}{8}
  = 17,187 \ \text{TFLOP/sec}.
\]
This number serves only as a theoretical upper bound, complementary to the vendor peak-performance figures shown in Table \ref{tab:tpu}.

\bibliographystyle{model1-num-names}

\bibliography{cas-refs}



\end{document}